\documentclass[a4paper,12pt]{article}
\usepackage{amssymb}
\usepackage{graphicx}
\usepackage{amsmath}
\usepackage{latexsym}
\usepackage{graphicx}
\usepackage{rotating}
\usepackage{amsmath}
\usepackage{amssymb}
\usepackage[bf,labelsep=period]{caption}
\usepackage{harvard}
\usepackage{dcolumn}
\usepackage{endnotes}
\usepackage[english]{babel}
\usepackage[dvips=true, pdftex=true, vtex=false]{geometry}
\usepackage{amsfonts}
\usepackage{lscape}
\usepackage{xcolor}
\usepackage{hyperref}

\newtheorem{theorem}{Theorem}
\newtheorem{lemma}{Lemma}

\newtheorem{assumption}{Assumption}

\newtheorem{result}{Result}

\renewcommand{\theequation}{\thesection.\arabic{equation}}

\numberwithin{equation}{section} 
\newcounter{rmk}
\newenvironment{rmk}[1][]{\refstepcounter{rmk}\par\medskip\noindent
  \textbf{Remark~\thermk.\ #1} \rmfamily} {\smallskip}
\newenvironment{remark}{\smallskip \begin{rmk}}{\hfill $\diamondsuit$ \end{rmk}  }

\input{tcilatex}

\begin{document}

\title{{\vspace{-5ex}Adaptive information-based methods for determining\vspace{-1ex} \\ the co-integration rank
in heteroskedastic VAR models\vspace{-1ex}
}}
\author{H.\ Peter Boswijk$^{a}$, Giuseppe Cavaliere$^{b,c,}$\thanks{Correspondence to: Giuseppe Cavaliere, University of Bologna, email \texttt{giuseppe.cavaliere@unibo.it}}, Luca De Angelis$^{b}$,
\and and A.\ M.\ Robert Taylor$^{c}$ \\
{\small {$^{a}$ Tinbergen Institute \& Amsterdam School of Economics, University of Amsterdam}}\\
$^{{b}}${\small {\ Department of Economics, University of Bologna}}\\
{\small {$^{c}$ Essex Business School, University of Essex}}}

\maketitle

\linespread{1.1} 

\begin{abstract}
\noindent Standard methods, such as sequential procedures based on
Johansen's (pseudo-)likelihood ratio (PLR) test, for determining
the co-integration rank of a vector autoregressive (VAR) system of
variables integrated of order one can be significantly affected,
even asymptotically, by unconditional heteroskedasticity
(non-stationary volatility) in the data. Known solutions to this
problem include wild bootstrap implementations of the PLR test or
the use of an information criterion, such as the BIC, to select
the co-integration rank. Although asymptotically valid in the
presence of heteroskedasticity, these methods can display very low
finite sample power under some patterns of non-stationary
volatility. In particular, they do not exploit potential
efficiency gains that could be realised in the presence of
non-stationary volatility by using adaptive inference methods.
Under the assumption of a known autoregressive lag length, Boswijk
and Zu (2022) develop adaptive PLR test based methods using a
non-parameteric estimate of the covariance matrix process. It is
well-known, however, that selecting an incorrect lag length can
significantly impact on the efficacy of both information criteria
and bootstrap PLR tests to determine co-integration rank in finite
samples.  We show that adaptive information criteria-based
approaches can be used to estimate the autoregressive lag order to
use in connection with bootstrap adaptive PLR tests, or to jointly
determine the co-integration rank and the VAR lag length and that
in both cases they are weakly consistent for these parameters in
the presence of non-stationary volatility provided standard
conditions hold on the penalty term. Monte Carlo simulations are
used to demonstrate the potential gains from using adaptive
methods and an empirical application to the U.S.\ term structure
is provided.

\bigskip


\bigskip

\noindent \textbf{J.E.L.\ Classifications}: C32, C14.

\bigskip

\end{abstract}

\thispagestyle{empty}

\linespread{1.3}

\section{Introduction}
\label{sec_intro}

It is well-known that standard methods for determining the
co-integration rank of vector autoregressive (VAR) systems of
variables integrated of order one are affected by the presence of
heteroskedasticity.  In particular, sequential procedures based on
(pseudo-) likelihood ratio [PLR] test as developed by Johansen
(1996) can be significantly over-sized, even in large samples,
when the volatility process displays non-stationary variation (so
called \emph{non-stationary unconditional volatility}) and,
moreover, the finite sample power of these tests can vary
enormously depending on the pattern of heteroskedasticity present;
see, in particular, Cavaliere, Rahbek and Taylor (2010). This is
an important issue in practice because time-varying behaviour in
unconditional volatility appears to be a common feature in many
key macroeconomic and financial time series; see, among many
others, McConnell and Perez Quiros (2000), Sensier and van Dijk
(2004), and Cavaliere and Taylor (2008); see also McAleer (2005, 2009), Asai et al., (2006) and  McAleer and Medeiros (2008).

In a series of recent papers, Cavaliere, Rahbek and Taylor (2010,
2014) show that a solution to the size problems induced by
non-stationary volatility is obtained by using wild bootstrap
based implementations of the standard PLR tests. In particular,
Cavaliere \emph{et al.}\ (2010) show that the sequential procedure
based on wild bootstrap PLR tests leads to consistent
co-integration rank determination in the presence of
non-stationary unconditional volatility.  As alternative solution
to the use of wild bootstrap PLR tests is considered by Cavaliere,
De Angelis, Rahbek and Taylor (2015, 2018) who show that methods
based on information criteria can also be used to consistently
determine the co-integration rank in the presence of
non-stationary volatility. In particular, they show that popular
information criteria such as the  Bayesian information criterion
[BIC] (Schwarz, 1978) and the Hannan-Quinn information criterion
[HQC] (Hannan and Quinn, 1979) provide a useful complement to the
wild bootstrap sequential procedures.

The wild bootstrap PLR tests are correctly sized in the presence
of non-stationary volatility and attain the same asymptotic local
power functions as infeasible size-corrected versions of the
standard PLR tests.  As such they can therefore display very low
power properties for some patterns of non-stationary volatility.
Indeed, other things equal, their asymptotic local power functions
are reduced, relative to the unconditionally homoskedastic case,
under non-stationary volatility.  Similarly, the ability of the
standard information criteria-based methods discussed above to
select the correct co-integration rank can also be greatly reduced
under non-stationary volatility.   In particular, none of these
methods exploits the potential efficiency gains that could be
provided by using inference methods which adapt to the volatility
process.  Adaptive methods, where the covariance matrix process is
estimated non-parametrically, have the potential to be
particularly useful in this context.

Under the assumption of a known autoregressive lag length, Boswijk
and Zu (2022) develop an procedure based on adaptive PLR tests for
determining the co-integration rank in possibly heteroskedastic
VAR models.  Specifically, they propose a procedure where the
volatility process is estimated using a non-parametric kernel
estimator, with this estimate then used in the adaptive PLR test
procedure. Under suitable conditions, they establish that the
non-parametric volatility estimator is consistent and that the
resulting adaptive PLR co-integration rank tests have the same
asymptotic local power functions as for infeasible tests based on
the assumption that the volatility process is known.  The
asymptotic null distribution of their proposed statistics are,
however, non-standard and depend on the realisation of the
volatility process.  As such, asymptotic $p$-values for the
adaptive PLR tests need to be obtained using bootstrap methods.

The assumption of a known of autoregressive lag order is
problematic in practice. It is well-known that an incorrect lag
length choice can significantly impact on the efficacy of both
information criteria and PLR tests, in particular where a lag
order smaller than the true order is used; see, among others,
Boswijk and Franses (1992), Cheung and Lai (1993), Haug (1996),
L\"{u}tkepohl and Saikkonen (1999), and Cavaliere \emph{et al.}\ (2018).
In practice the autoregressive lag length will need to
be 
estimated along with the co-integration rank.  To that end, the
practitioner can use either a sequential procedure, where the lag
length is consistently estimated in a first step and then
subsequently employed in the second step in a procedure such as
either
the adaptive PLR test approach of Boswijk and Zu (2022)
or an information criterion
for determining the co-integration rank, or a joint information
criteria-based approach can be used whereby the lag length and
co-integration rank are determined simultaneously. Cavaliere
\emph{et al.}\ (2018) show that both joint and sequential procedures based on standard
information criteria consistently determine both the lag length
and the co-integration rank in the presence of non-stationary
unconditional volatility, provided standard conditions hold on the
penalty term.  They also show the asymptotic validity of a
sequential procedure based on wild bootstrap PLR tests with the
autoregressive lag length chosen by an information criterion.

The contribution of this paper is to develop adaptive information
criteria methods, based around a (non-parametric) estimation of
the volatility process, for jointly
selecting the co-integration
rank and autoregressive lag order.
We show that these adaptive
information criteria-based methods are weakly consistent for the
co-integration rank and autoregressive lag order under the
precisely the same conditions on the penalty function are as
required for the consistency of standard (non-adaptive)
information criteria under non-stationary volatility of the form
considered in this paper.
We also establish the asymptotic validity of a sequential procedure
selecting the autoregressive lag length by an adaptive information criterion [ALS-IC]
in the first step and then determining the co-integration rank using again an ALS-IC
in the second step based on the first step
estimate of the lag length.
Because the co-integration rank is
determined by minimising an adaptive information criterion 
over all possible values of the co-integration rank from zero up
to the dimension of the system, the practitioner does not
therefore need to obtain $p$-values by bootstrap methods, making
the procedure considerably less time consuming than the Boswijk
and Zu (2022) procedure based on adaptive PLR tests. We also
establish the asymptotic validity of a sequential procedure
selecting the autoregressive lag length by an ALS-IC in the first
step and then using the adaptive PLR test-based approach of
Boswijk and Zu (2022) in the second step based on the first step
estimate of the lag length.



The remainder of the paper is organised as follows. Section
\ref{sec_model} details our reference heteroskedastic
co-integrated VAR model. Section \ref{sec_ALS} outlines adaptive information criteria-based
methods for determining the co-integration rank and the
autoregressive lag length.  The large sample properties of these
procedures are detailed in Section \ref{sec_asy}. Monte Carlo
simulation experiments reported in Section \ref{sec_MC} are used
to explore the finite sample performance of the ALS-IC methods
relative to standard methods such as those based on standard
information criteria-based procedures.   These results highlight
the potential gains that can be achieved by using adaptive
methods. Section \ref{sec_emp} provides an empirical application
of the methods discussed in this paper to the term structure of
interest rates in the US.  Section \ref{sec_conc} concludes.
Proofs of our main results 
 are contained
in the 
 Appendix \ref{app}.

\section{The Heteroskedastic Co-integrated VAR Model}
\label{sec_model}

Consider the $p$-dimensional process $\{ X_{t} \} $ which satisfies the $k$%
-th order reduced rank VAR model:
\begin{equation}
\Delta X_{t}=\alpha \beta^{\prime}X_{t-1}+\sum_{i=1}^{k-1}\Gamma_{i}\Delta
X_{t-i}+\alpha \rho^{\prime}D_{t}+\phi d_{t}+\varepsilon_{t}, \; \; t=1,\ldots,T ,
\label{1}
\end{equation}
where $X_t := ( X_{1t} ,\ldots, X_{pt})^\prime $ and the initial values, $%
X_{1-k},\ldots,X_0 $, are taken to be fixed in the statistical analysis. Let $%
k_0$ denote the true value of the autoregressive lag length $k$ in \eqref{1}%
. In the context of \eqref{1} we assume that the standard
`I$(1,r_0)$ conditions' hold, where $r_0 \in \{ 0, \ldots, p \} $
denotes the true co-integration rank of the system (see also
Cavaliere, Rahbek and Taylor, 2012); that is, the characteristic
polynomial associated with \eqref{1} has $p-r_0$ roots equal to 1
with all other roots lying outside the unit circle, and where
$\alpha$ and $\beta$ have full column rank $r_0$.

The deterministic variables in \eqref{1} are taken to satisfy one of the
following cases (see, e.g., Johansen, 1996): (i) $D_{t}=0$, $d_{t}=0$ (no
deterministic); (ii) $D_{t}=1$, $d_{t}=0$ (restricted constant); or (iii) $%
D_{t}=t$, $d_{t}=1$ (restricted linear trend). 

The innovation process $\varepsilon_t := (\varepsilon_{1t} ,\ldots,
\varepsilon_{pt} )^\prime$ in \eqref{1} is taken to satisfy the
following set of conditions collectively labelled Assumption
\ref{ass1}.

\begin{assumption} \label{ass1} The innovations $%
\{\varepsilon_{t}\}$ are defined as $\varepsilon_{t}:=\sigma _{t}z_{t}$,
where $\sigma _{t}$ is non-stochastic
and satisfies $\sigma_{t}:=\sigma \left(t/T\right) $ for all $t=1,\ldots,T$,
where $\sigma \left(\cdot \right) \in $\textit{$\mathcal{D}_{\mathbb{R}%
^{p\times p}}[0,1]$}, with
$\mathcal{D%
}_{\mathbb{R}^{m\times n}}[0,1]$ used to denote the space of
$m\times n$ matrices of c\`{a}dl\`{a}g functions on $[0,1]$
{equipped with the Skorokhod metric,} and where
$\sigma \left( u\right)$ is non-singular for all $u \in \lbrack 0,1]$ and continuous in $u \in \lbrack 0,1]$;
$z_{t}$ is an i.i.d.\ sequence with $E(z_t)=0$ and $E(z_t z_t^\prime)=I_p$.
\end{assumption}

\medskip

\begin{remark} Assumption \ref{ass1} implies that $E(\varepsilon_{t})=0$ and
that $ \varepsilon_{t} $ has the time-varying unconditional
variance matrix
$\Sigma_{t}:=E(\varepsilon_{t}\varepsilon_{t}^\prime) = \sigma_{t}
\sigma_{t}^\prime >0$.  In what follows, $\sigma_{t}$ will be
referred to as the \emph{volatility matrix} of $\varepsilon_{t}$.
Elements of Assumption \ref{ass1} have previously been employed
by, \emph{inter alia}, Cavaliere \emph{et al.}\ (2010), Boswijk,
Cavaliere, Rahbek and Taylor (2016), Cavaliere \emph{et al.}\
(2018) and Boswijk and Zu (2022). In particular, Assumption 1
allows for a countable number of discontinuities in $\sigma
\left(\cdot \right)$ therefore allowing for a wide class of
potential models for the time-varying behaviour of the
unconditional variance matrix of $\varepsilon_{t}$.  As discussed
in Boswijk and Zu (2022), the continuity assumption on $\sigma
\left( \cdot \right)$ is made so that $\sigma \left(\cdot \right)$
can be consistently estimated. This assumption is not restrictive
in practice however because one can always approximate
discontinuities in $\sigma \left( \cdot \right)$ arbitrarily well
using smooth transition functions. Moreover, one could relax this
assumption by assuming that $ \sigma ( \cdot ) $ is a piecewise
Lipschitz-continuous function; see Xu and Phillips (2008).
\end{remark}

\begin{remark} In order to simplify our presentation, Assumption \ref{ass1} rules out the possibility of conditional
heteroskedasticity in $z_{t}$.  We do so because adaptive
estimation can only lead to efficiency gains over standard
estimation in cases where $ \sigma( u ) $ varies across $ u$ which
can only happen where non-stationary volatility is present.
Conditional heteroskedasticity of the form considered in
Assumption 2(b) of Boswijk \emph{et al.}\ (2016),
cannot induce time-variation in $ \sigma( u ) $ and so it is
irrelevant so far as adaptive estimation is concerned. It is
straightforward, however, to show that the large sample results
given in this paper remain valid if we allow for conditional
heteroskedasticity in $ z_t$ of the form considered in Assumption
2(b) of Boswijk \emph{et al.}\ (2016).
\end{remark}

\section{Adaptive Information Criteria} 
\label{sec_ALS}

In this section we discuss adaptive information-based methods for
determining the co-integration rank and the autoregressive lag
length in the context of \eqref{1}. In particular, we first derive
the log-likelihood function in Section \ref{sec_LL} and the
nonparametric estimator of the volatility matrix in Section
\ref{sec_vol}.  We then outline the adaptive information criterion
for the joint determination of the co-integration rank and the lag
length in Section \ref{sec_IC} and we discuss how to sequentially
estimate the lag length and the co-integration rank using adaptive
methods in Section \ref{sec_Seq}.

\subsection{The Likelihood Function}
\label{sec_LL}

Define $\Psi:=[\Gamma_1 : \ldots : \Gamma_{k-1}]$ and $Z_t^{(k)}:=(\Delta X_{t-1}^\prime, \ldots, \Delta X_{t-k+1}^\prime)^\prime$, such that the model in \eqref{1} with no deterministic components (case (i)) can be rewritten more compactly as
\begin{equation}
\Delta X_t = \alpha \beta X_{t-1} + \Psi Z_t^{(k)}+\varepsilon_t .
\label{2}
\end{equation}

Suppose for the present that $\{\sigma_t\}$ is known, and that
$z_t$ is Gaussian; i.e., $z_t \sim \mathrm{i.i.d.}\ N(0, I_p)$.  Then under Assumption 1 we have that $\varepsilon_t | \mathcal{F}_{t-1} \sim N(0, \Sigma_{t})$, where 
$\mathcal{F}_{t-1}:=\{X_{t-1}, \ldots, X_1, X_0, \linebreak \ldots, X_{1-k} \}$,
and the log-likelihood function is given by (see Boswijk and Zu, 2022):
\begin{eqnarray}
\nonumber
\ell_T (\alpha, \beta, \Psi) &=& -\frac{Tp}{2} \log 2\pi - \frac{1}{2} \sum_{t=1}^T \log |\Sigma_{t}| \\
& &  - \frac{1}{2} \sum_{t=1}^T (\Delta X_t - \alpha \beta^\prime X_{t-1} - \Psi Z_t^{(k)} )^\prime \Sigma_{t}^{-1} (\Delta X_t - \alpha \beta^\prime X_{t-1} - \Psi Z_t^{(k)}).
\label{LL}
\end{eqnarray}

Maximum likelihood estimation of the parameters $(\alpha, \beta,
\Psi)$ can be achieved by using the so-called \emph{generalised
reduced rank regression} procedure (Boswijk, 1995; Hansen, 2002,
2003), which uses a switching algorithm in order to circumvent the
issue of the lack of a closed-form expression for the maximum
likelihood estimator (MLE). In particular, because the MLE of
$(\alpha, \Psi)$ for fixed $\beta$ and the MLE of $\beta$ for
fixed $(\alpha, \Psi)$ have closed-form expressions, the
maximisation of \eqref{LL} can be achieved, starting from an
initial guess, by switching between maximisation over $(\alpha,
\Psi)$ and $\beta$; see Boswijk and Zu (2022) for further details.

\subsection{Volatility Estimation}
\label{sec_vol}

In this paper we focus on the two-sided smoothing nonparametric
estimator of the volatility matrix adopted by Boswijk and Zu
(2022). This estimator is a multivariate extension of Hansen
(1995)'s nonparametric volatility filter based on leads and lags
of the outer product of the residual vector.  A similar approach
to adaptive estimation has also been considered by Xu and Phillips
(2008) and Patilea and Ra\"{i}ssi (2012), among others.

Let $K(\cdot)$ denote some kernel function and define $K_h (x) :=
K(x/h)/h$ with $h>0$ a {\it window width}. The kernel estimator
for $\Sigma_t$ that we will consider is then defined as,
\begin{equation}
\hat{\Sigma}_t := \frac{\sum_{s=1}^{T}
K_h\left(\frac{t-s}{T}\right) \hat{e}_{s}\hat{e}_{s}^\prime}
{\sum_{s=1}^{T} K_h\left(\frac{t-s}{T}\right)}, \label{vol}
\end{equation}
where $\hat{e}_t$ is the residual vector obtained by estimating an
unrestricted VAR model of order $K$ in the levels of $X_t$, i.e.\
$\hat{e}_t = X_t - \sum_{i=1}^K \hat{A}_i X_{t-i}$, where $A_i$,
$i=1,\ldots,K$, are $p \times p$ coefficient matrices. The value $K $
denotes the maximum autoregressive lag order we will allow for
which, unless otherwise stated, is assumed in the following to be
at least as large as the true lag order, $k_0$ in \eqref{1}.

The kernel function in \eqref{vol} is implemented with two-sided
smoothing, so that  $\hat{\Sigma}_t$ is based on leads and lags of
$\hat{e}_t \hat{e}_t^\prime$, as outlined in Assumption 3 in
Boswijk and Zu (2022). In their Lemma 2, Boswijk and Zu (2022)
show that the volatility matrix process implied by the $T$
nonparametrically estimated covariance matrices is uniformly
consistent over the compact interval $[0,1]$, which, in turn,
implies uniform consistency of the nonparametric estimator
$\hat{\Sigma}_t$ in \eqref{vol} over $t=1,\ldots,T$. Therefore, these
consistent estimators can be used to replace $\Sigma_t$ in the
log-likelihood function in \eqref{LL}, thereby allowing for a
feasible version of the generalised reduced rank regression
procedure and the computation of the adaptive information criteria
and the adaptive bootstrap PLR tests.

In implementing the nonparametric estimator of $ \Sigma_t $ in
\eqref{vol}, we will select the window width $h$ by minimising the
quantity
\begin{equation*}
\tilde{C}_T(h) := \sum_{t=1}^T || \hat{\Sigma}_t^{-t}(h) -
\hat{e}_{t}\hat{e}_{t}^\prime ||^2,
\end{equation*}
where $||\cdot||$ denotes the Euclidean matrix norm, and where
$\hat{\Sigma}_t^{-t}(h)$ is given by \eqref{vol}, but with $K(0)$
replaced by 0, so that $\hat{e}_{t}\hat{e}_{t}^\prime$ does not
enter the expression for $\hat{\Sigma}_t^{-t}(h)$. This
leave-one-out cross-validation technique is implemented in Boswijk
and Zu (2016) and Patilea and Ra\"{i}ssi (2012), and satisfies the
requirement that $h$ decreases with the sample size at a certain
rate; see Lemma 2 of Boswijk and Zu (2022) and Section
\ref{sec_asy} below.

\subsection{Joint Determination of the Lag Length and Co-integration Rank}
\label{sec_IC}

The maximised pseudo log-likelihood function \eqref{LL} associated with \eqref{2}
under lag order $k$ and co-integration rank $r$, say $\hat{\ell}_T^{(k,r)} (\alpha, \beta, \Psi)$, in conjunction
with the volatility estimator in \eqref{vol} substituted for $\Sigma_t$ in \eqref{LL} can then be used to construct a feasible \emph{adaptive} information criterion of the following generic form
\begin{equation}
\text{ALS-IC}(k,r) := -2 \hat{\ell}_T^{(k,r)} (\alpha, \beta, \Psi) + c_T \pi(k,r),
\label{ICIC}
\end{equation}
where the term $c_T$ may depend on the sample size $T$ (see below)
and where $\pi(k,r)$ denotes the number of parameters in the
estimated model.\footnote{The number of parameters which defines
the penalty term in \eqref{ICIC} depends on the deterministic
components included in the model \eqref{1} as follows:
(i) in the case of no deterministic component ($D_t=0$, $d_t=0$ in %
\ref{1}), $\pi(k,r) = r(2p-r)+p^2(k-1)$; (ii) for the
restricted constant case ($D_t=1$, $d_t=0$ in \ref{1}), $\pi(k,r)
= r(2p-r+1)+p^2(k-1)$, and (iii) for the case of a restricted trend ($%
D_t=1$, $d_t=1$ in \ref{1}), $\pi(k,r) =
r(2p-r+1)+p+p^2(k-1)$.
}
The autoregressive lag order and the co-integration rank can then
be jointly estimated by minimising the information criterion in
\eqref{ICIC} jointly over both all possible lag lengths, $k =
1,\ldots,K$, and over all possible co-integration ranks, $r =
0,\ldots,p $; that is,
\begin{equation*}
( \tilde{k}_{\mathrm{ALS}\text{-}\mathrm{IC}}, \tilde{r}_{\mathrm{ALS}\text{-}\mathrm{IC}} ) := \operatornamewithlimits{%
\text{arg} \min}_{r=0,\ldots,p; k=1,\ldots,K} \text{ALS-IC}(k,r).
\label{hat}
\end{equation*}

Different values of the coefficient $c_T$ yield different adaptive information criteria. 
In the standard (non-adaptive) case, which can be obtained as a
special case of the adaptive information criterion in \eqref{ICIC}
by restricting $\Sigma_t = I_p$ in the likelihood function
\eqref{LL}, the most widely used information criteria are the
Akaike information criterion [AIC] (Akaike, 1974), the Bayes
information criterion [BIC] (Schwarz, 1978), and the Hannan-Quinn
information criterion [HQC] (Hannan and Quinn, 1979), which obtain
setting $c_T=2$, $\log T$, and $2 \log \log T$, respectively.  We
will denote the generic standard information criterion in this
case as $\text{IC}(k,r)$ and the resulting estimate in \eqref{hat}
as $ ( \tilde{k}_{\rm IC}, \tilde{r}_{\rm IC} ) $. In the context
of \eqref{ICIC}, we will refer to the adaptive information
criteria based on the AIC, BIC and HQ choices of $c_T$ as ALS-AIC,
ALS-BIC, and ALS-HQC, respectively.

\subsection{Sequential Determination of the Lag Length and Co-integration Rank}
\label{sec_Seq}

Because the lag length $k$ in \eqref{1} is in general unknown and
needs to be estimated prior to estimating the co-integration rank,
practitioners often use a two-step procedure, whereby the
autoregressive lag length is estimated in the first step and then
subsequently employed as if it were the known lag length in a
second step for determining the co-integration rank, such as a
sequential procedure based on PLR tests or an information
criterion. In particular, L\"{u}tkepohl and Saikkonen (1999) and
Nielsen (2006), \emph{inter alia}, show that
the lag length in nonstationary VAR models can be consistently estimated from the levels of the data using an information criterion.
Therefore, the lag length could be selected in the first step of the sequential procedure according to a (standard) information criterion where we do not impose a reduced rank structure on $\Pi:=\alpha\beta^\prime$ in \eqref{1}, that is by imposing $r=p$; see, among others, Cavaliere \emph{et al.}\ (2018).

As for the joint determination of the lag length and the co-integration rank
considered in Section \ref{sec_IC}, an adaptive version of the
information criterion for determining the lag length can also be
considered.  In particular, the lag length may be selected using
an adaptive information criterion of the generic form
\begin{equation}
\text{ALS-IC}(k,p) := -2 \hat{\ell}_T^{(k,p)}(\Pi, I_p, \Psi) + c_T \pi(k,p),
\label{IC(k)}
\end{equation}
 where $\hat{\ell}_T^{(k,p)}(\Pi, I_p, \Psi)$ is the maximised pseudo likelihood \eqref{LL} associated with \eqref{2} where
we do not impose a reduced rank structure on $\Pi=\alpha
\beta^\prime$ under lag length $k$ and $\Sigma_t$ in \eqref{LL} is
substituted with the volatility estimator in \eqref{vol}. Again,
the choice of the $c_T$ term identifies different information
criteria as outlined above and, in this case, $\pi(k,p)=p(pk+i)$
with $i = 0$ when no deterministic component is involved, $i = 1$
in the case of restricted constant, and $i = 2$ for the restricted
trend. The resulting adaptive information criterion-based lag
length estimator is then given by
\begin{equation*}
\hat{k}_{\text{ALS-IC}} := \operatornamewithlimits{%
\text{arg} \min}_{k=1,\ldots,K} \text{ALS-IC}(k,p).
\label{hat_k}
\end{equation*}

We note again that the generic standard information criterion,
which we will denote by $\text{IC}(k,p)$, can be obtained as a
special case of the adaptive information criterion in
\eqref{IC(k)}, by restricting $\Sigma_t = I_p$ in the likelihood
function \eqref{LL}, with the resulting lag length estimator
denoted by $\hat{k}_{\text{IC}}$. In the simulation experiments
discussed in Section \ref{sec_MC}, we will consider both the
standard and the adaptive versions of the information criterion
for determining the lag length in the first step of the two step
sequential procedure.  The selected lag length, either
$\hat{k}_{\text{IC}}$ or $\hat{k}_{\text{ALS-IC}}$ generically
denoted by $\hat{k}$ for the remainder of this section,  is then
used as if it were the true lag length in the second step for
determining the co-integration rank.  The second step could be
based on either the sequential procedure of Boswijk and Zu (2022)
based on adaptive bootstrap PLR tests or an adaptive information
criterion for selecting the co-integration rank.  We now outline
these two possibilities.

\smallskip

\emph{The adaptive PLR test-based procedure of Boswijk and Zu (2022).}
Boswijk and Zu (2022) introduce the adaptive PLR statistic  for testing the null hypothesis that
the true co-integration rank is (no more than) $r$, $0 \leq r \leq p-1 $,
\begin{equation}
Q_{r,k,T} := -2\left[\hat{\ell}_T^{(k,r)} (\alpha, \beta, \Psi) - \hat{\ell}_T^{(k,p)} (\Pi, I_p, \Psi)  \right] = \sum_{t=1}^{T} \left(\hat{\varepsilon}_{r,t}^{\prime} \hat{\Sigma}_t^{-1} \hat{\varepsilon}_{r,t} - \hat{\varepsilon}_{p,t}^{\prime} \hat{\Sigma}_t^{-1} \hat{\varepsilon}_{p,t}
\right),
\label{PLR0}
\end{equation}
where $\hat{\varepsilon}_{r,t}$ and $\hat{\varepsilon}_{p,t}$ denote the residuals from the restricted and unrestricted VAR model in \eqref{1}, respectively.
For the case where the autoregressive lag length is known ($k=k_0$), they demonstrate that the limiting distribution of \eqref{PLR0}
depends on the unknown volatility process.
Consequently, bootstrap methods are required to approximate the
critical values from this distribution. In order to do so, a
bootstrap sample $\{X_{r,t}^\ast \}_{t=1}^T$ is generated
recursively from
\begin{equation}
\Delta X_{r,t}^\ast = \hat{\alpha}^{(r)} \hat{\beta}^{(r)\prime} X_{r,t-1}^\ast
+ \sum_{i=1}^{k-1}  \hat{\Gamma}_i^{(r)} \Delta X_{r,t-i}^\ast + \varepsilon_{r,t}^\ast, \ \ \ t=1, \ldots, T,
\label{BSsample}
\end{equation}
initialised at $X_{r,j}^\ast = X_{j}$, for $j=1-k, \ldots, 0$, where
$\hat{\alpha}^{(r)}$, $\hat{\beta}^{(r)}$, and
$\hat{\Gamma}_i^{(r)}$ are the estimated parameter matrices from
the model \eqref{1} obtained using conventional reduced rank
regression under the rank $r$ imposed by the null hypothesis. The
adaptive PLR test statistic based on the bootstrap sample is then
computed as
\begin{equation}
Q_{r,k,T}^\ast := \sum_{t=1}^{T} \left(\hat{\varepsilon}_{r,t}^{\ast\prime} \hat{\Sigma}_t^{-1} \hat{\varepsilon}_{r,t}^\ast -
\hat{\varepsilon}_{p,t}^{\ast\prime} \hat{\Sigma}_t^{-1} \hat{\varepsilon}_{p,t}^\ast \right),
\label{PLR}
\end{equation}
where $\hat{\varepsilon}_{r,t}^\ast$ and $\hat{\varepsilon}_{p,t}^\ast$ denote the (bootstrap) residuals from the restricted and unrestricted models, respectively.
Following Boswijk and Zu (2022), we consider the following two
bootstrap implementations: (i) the variance bootstrap,
$\varepsilon_{r,t}^\ast := \hat{\Sigma}_t^{1/2} z_t^\ast$, where
$\hat{\Sigma}_t^{1/2}$ is any square root of $\hat{\Sigma}_t$ and
$z_t^\ast \sim \mathrm{i.i.d.} N(0, I_p)$; (ii) the wild bootstrap,
$\varepsilon_{r,t}^\ast := \hat{\varepsilon}_{r,t} w_t^\ast$,
where 
$w_t^\ast$ is a scalar i.i.d.\ N(0,1)
sequence; see Section 4.2 of Boswijk and Zu (2022) for more details.
As is typically done in practice, the unknown lag length $k$ in \eqref{BSsample} and \eqref{PLR} is replaced by the lag length estimated in the first step of the sequential procedure, 
say $\hat{k}$, in order to compute the bootstrap statistic $Q_{r,\hat{k},T}^\ast$ using the bootstrap sample in \eqref{BSsample} based on $\hat{k}$.
The corresponding $p$-value is then computed as $p^\ast_{r,\hat{k},T}:=1-G^\ast_{r,\hat{k},T}(Q_{r,\hat{k},T}^\ast)$, where $G^\ast_{r,\hat{k},T}(\cdot)$ denotes the conditional (on the original data) cdf of $Q_{r,\hat{k},T}^\ast$.
Starting from $r=0$, the bootstrap algorithm is repeated as long as $p^\ast_{r,\hat{k},T}$ exceeds the significance level $\eta$, thus yielding $\hat{r}^\ast(\hat{k})=r$. If the null is not rejected for $r=p-1$, then $\hat{r}^\ast(\hat{k})=p$.

The asymptotic validity of the two bootstrap procedures outlined above
is established in Theorem 3 of Boswijk and Zu  (2022) with the
implication that, for the case where the autoregressive lag length
is known ($k=k_0$),
the variance and wild bootstrap adaptive PLR test-based procedures,  $\hat{r}^\ast(k_0)$, 
 are asymptotically accurately capped estimator
of the co-integration rank
$r_0$.\footnote{The sequential rank determination procedure of Johansen (1996) is \emph{asymptotically accurately capped} in that if each PLR (or bootstrap PLR) test in the
sequence is run with nominal (asymptotic) significance level $
\eta $, then the limiting probability of selecting a rank smaller
than, equal to, and greater than the true rank will be 0, $1-
\eta $ and $\eta $, respectively, when $r_0 < p $ and 0, 1 and 0,
respectively, when $r_0 = p $.}
In Section \ref{sec_asy} we will generalise these results to the case where the lag length is unknown and estimated in the first step of the sequential procedure.

\smallskip

\emph{The adaptive IC-based procedure.}
Alternatively, to determine the co-integration rank in the second step of a sequential procedure based on ALS-IC, $k$ in the generic form \eqref{ICIC} can be replaced
by the lag length estimated in the first step, 
thus yielding
\begin{equation}
\text{ALS-IC}(\hat{k},r) := -2 \hat{\ell}_T^{(\hat{k},r)}(\alpha, \beta, \Psi) + c_T \pi(\hat{k},p).
\label{IC(r)}
\end{equation}
The resulting adaptive information criterion-based co-integration rank estimator is then given by
\begin{equation*}
\hat{r}_{\text{ALS-IC}}(\hat{k}) := \operatornamewithlimits{%
\text{arg} \min}_{r=0,\ldots,p} \text{ALS-IC}(\hat{k},p).
\label{hat_r}
\end{equation*}

\section{Asymptotic Analysis}
\label{sec_asy}

In this section we establish the large sample properties of the
adaptive methods for determining the co-integration rank and
autoregressive lag length outlined in Sections \ref{sec_IC} and
\ref{sec_Seq}.

Lemma 2 of Boswijk and Zu (2022) establishes that the
nonparameteric estimate of the volatility matrix process defined
as $\hat{\Sigma}_T(u):=\sum_{t=1}^T \hat{\Sigma}_t
1_{[(t-1)/T,t/T]} (u)$ is uniformly consistent over the compact
interval $[0,1]$. This result is a basic building block needed to
demonstrate weak consistency\footnote{%
An estimator $T_{n}$ is defined to be \emph{weakly} consistent if it
converges in \emph{probability} to the true value of the unknown parameter $%
\theta$; that is, $T_{n} \overset{p}{\rightarrow } \theta$.} for
the adaptive information criteria in \eqref{ICIC} and
\eqref{IC(k)} and so for completeness we first reproduce that
result below as Result 1.

\medskip

\begin{result}
\noindent Let $\{X_{t}\}$ be generated as in \eqref{1} with the parameters satisfying the I(1, $r_{0}$) conditions and let Assumption 1 hold, and let $K$ be a bounded non-negative function defined on $\mathbb{R}$ which satisfies $\int_{-\infty}^{\infty}K(x)dx=1$, $0<\int_{-\infty}^{0}K(x)dx<1$ and $0<\int_{0}^{\infty}K(x)dx<1$.
Then, if $T\to \infty$, $h \to 0$ and $Th^2 \to \infty$, it holds that
$$\operatornamewithlimits{\sup}_{u\in [0,1]} || \hat{\Sigma}_T(u) - {\Sigma}(u)  || \overset{p}{\to}0 , $$ 
where ${\Sigma}(u):=\sigma(u) \sigma(u)^\prime$ is the true variance matrix process.
\end{result}

\medskip

Using Result 1, we first show in Lemma \ref{Th_joint_r} that the
adaptive information criterion in \eqref{ICIC} is weakly
consistent for the co-integration rank, regardless of the
autoregressive lag length used, provided standard conditions hold
on the penalty term, $ c_T$.  Then second in Lemma
\ref{Th_joint_k} we show that for the true co-integration rank, $
r_0$, the adaptive information criterion in \eqref{ICIC} is weakly
consistent for the autoregressive lag length.

\medskip

\begin{lemma}  Let the conditions of Result 1 hold.  Then, for any
$0 < k \le K$, it holds that, as $T \rightarrow \infty$:

\begin{description}
\item (i) for $r>r_{0}$,
$\Pr\left(\mathrm{ALS}\text{-}\mathrm{IC}(k,r)>\mathrm{ALS}\text{-}\mathrm{IC}(k,r_{0})\right)
\rightarrow 1$, provided $c_{T} \rightarrow \infty$;

\item (ii) for $r<r_{0}$, $\Pr\left( \mathrm{ALS}\text{-}\mathrm{IC}(k,r)> \mathrm{ALS}\text{-}\mathrm{IC}
(k,r_{0})\right) \rightarrow 1$, provided $c_{T}/T \rightarrow 0$.
\end{description}

\label{Th_joint_r}
\end{lemma}

\medskip

\begin{lemma}
Let the conditions of Result 1 hold. Then it holds that, as $T
\rightarrow \infty$:

\begin{description}
\item (i) for any $k$ such that $k_{0} < k \le K$, $\Pr\left(\mathrm{ALS}\text{-}\mathrm{IC}
(k,r_{0})>\mathrm{ALS}\text{-}\mathrm{IC}(k_0,r_{0})\right) \rightarrow 1$, provided
$c_{T} \rightarrow \infty$;

\item (ii) for any $k$ such that $0<k<k_{0}$, $\Pr\left(
\mathrm{ALS}\text{-}\mathrm{IC}(k,r_0)> \mathrm{ALS}\text{-}\mathrm{IC}(k_0,r_{0})\right) \rightarrow
1$, provided $c_{T}/T \rightarrow 0 $.
\end{description}

\label{Th_joint_k}
\end{lemma}

\begin{remark} The results in Lemma \ref{Th_joint_r} imply that, provided
the standard condition that $\frac{c_T}{T}+\frac{1}{c_T} \,
{\rightarrow} \, 0 $, as $T \rightarrow \infty $, holds on the
penalty term, $ c_T$,  then for any lag length $k=1,\ldots, K$, the
adaptive information criterion-based estimator of the
co-integration rank is weakly consistent for the true
co-integration rank, $r_0$. The results in Lemma \ref{Th_joint_k}
imply that, under the same conditions on $ c_T$, the adaptive
information criterion-based estimator
of the lag length, computed by imposing the true co-integration rank, i.e.\ $r=r_0$ in %
\eqref{1},
is a weakly consistent estimator for the true lag order $k_0$.
Consequently, in each case, the use of either the ALS-BIC or
ALS-HQC, but not the ALS-AIC penalty, will yield weakly consistent
estimates.  Cavaliere \emph{et al.}\ (2018) demonstrate that
analogous results hold, with the same condition on $c_T$, for the
corresponding non-adaptive information criterion-based estimators.
\end{remark}

\medskip

Using the results in Lemmas \ref{Th_joint_r} and \ref{Th_joint_k},
we are now in a position to establish the weak consistency of the
joint procedure.  This is now given in Theorem \ref{Th_joint}.

\begin{theorem}
Let the conditions of Result 1 hold. Then it
holds that $( \tilde{k}_{{\mathrm{ALS}\text{-}\mathrm{IC}}}, \tilde{r}_{{\mathrm{ALS}\text{-}\mathrm{IC}}}) \overset{p}{\to }%
\left( {k}_{0}, {r}_{0} \right) $, provided $c_T$ in \eqref{ICIC}
satisfies the condition that $\frac{c_T}{T}+\frac{1}{c_T} {\rightarrow} 0$
as $T \rightarrow \infty$.
\label{Th_joint}
\end{theorem}

\begin{remark} An immediate consequence of the result in Theorem %
\ref{Th_joint} is that the resulting ALS-BIC-based and
ALS-HQC-based estimators are weakly consistent for both the
co-integration rank and autoregressive lag length, but that the
corresponding ALS-AIC-based estimator is not.
\end{remark}

\medskip

To conclude this section we now detail the large sample behaviour
of the two-step sequential procedures outlined in Section
\ref{sec_Seq} where in the first step we select the autoregressive
lag and then in the second step an adaptive procedure based on
this estimated lag length is used to determine the co-integration
rank.

First, in Lemma \ref{Th_k}, we generalise the results in Lemma 3
of Cavaliere \emph{et al.}\ (2018), which show the sufficient
conditions on the term $c_T$ that ensure weak consistency for an
information criterion of the form given in \eqref{IC(k)}, to the
case of its adaptive analogue, ALS-IC($k,p$). In particular, we
derive the conditions under which minimising an adaptive
information criterion consistently selects the true lag order,
$k_0$, in the first step when we do not impose a reduced rank
structure, so that we set $r=p$.

\begin{lemma}
Let the conditions of Result 1 hold. Then, for any $0 < k \le K,$
it holds that, as $T \rightarrow \infty$:

\begin{description}
\item (i) for $k>k_0$, $\Pr\left(\mathrm{ALS}\text{-}\mathrm{IC}%
(k,p)>\mathrm{ALS}\text{-}\mathrm{IC}(k_0,p)\right) \rightarrow 1$, provided $c_{T}
\rightarrow \infty$;

\item (ii) for $k < k_0$, $\Pr\left( \mathrm{ALS}\text{-}\mathrm{IC}(k,p)>
\mathrm{ALS}\text{-}\mathrm{IC}(k_0,p)\right) \rightarrow 1$, provided $c_{T}/T
\rightarrow 0 $.
\end{description}

\label{Th_k}
\end{lemma}

The results in Lemma \ref{Th_k} imply that
$\hat{k}_{\text{ALS-IC}} \overset{p}{\to} k_0$, again provided
$\frac{c_T}{T}+\frac{1}{c_T} {\rightarrow} 0$, as $ T \rightarrow
\infty $.
Using the results in Lemmas \ref{Th_joint_r} and \ref{Th_k}, we are now in a position in
Theorem \ref{Th_PLR} to establish the large sample properties of the 
bootstrap adaptive PLR test-based estimator of the co-integration rank using the lag length estimated by an 
information criterion as in \eqref{IC(k)} at the first step,  $\hat{r}^\ast(\hat{k}_{\text{ALS-IC}})$.

\begin{theorem}
Let the conditions in Result 1 hold. Then, provided $c_T$ in
\eqref{IC(k)} is such that $\frac{c_T}{T}+\frac{1}{c_T}
{\rightarrow} 0$ as $T \rightarrow \infty $, the variance and the
wild bootstrap PLR-tests satisfy:

\begin{description}
\item (i) $\lim_{T\rightarrow\infty}\Pr ( \hat{r}^\ast(\hat{k}_{\mathrm{ALS}\text{-}\mathrm{IC}}) =r ) =0$ for all $r=0,1,\ldots,r_{0}-1$%
;

\item (ii) $\lim_{T\rightarrow\infty}\Pr ( \hat{r}^*(\hat{k}_{\mathrm{ALS}\text{-}\mathrm{IC}}) =r_{0} ) =1-\eta\cdot\mathbb{I}(r_{0}<p)$, and $\underset{%
T\rightarrow \infty}{\lim}\underset{r\in\{r_{0}+1,\ldots,p\}}{\sup}\Pr( \hat{r}^*(\hat{k}_{\mathrm{ALS}\text{-}\mathrm{IC}}) =r ) \leq\eta$.
\label{Th_PLR} \end{description}
\end{theorem}

\begin{remark}
The results in Theorem \ref{Th_PLR} show that, provided the
information criterion used in the first step of the sequential
procedure is a consistent lag length estimator, that is
$\frac{c_T}{T}+\frac{1}{c_T} {\rightarrow} 0$, as $ T \rightarrow
\infty $, the bootstrap adaptive PLR test-based procedure is an
asymptotically accurately capped estimator
of the true co-integration rank, $r_0$.
\end{remark}

\begin{remark} The results in Theorem
\ref{Th_PLR} can also be shown to hold (under the same conditions)
for any consistent lag length estimator obtained in the first
step. Therefore, the consistency result in Theorem \ref{Th_PLR}
will also hold for variance and wild bootstrap adaptive PLR tests
when a standard information criterion, such either BIC($k,p$) or
HQC($k,p$), is used to select the lag length at the first step.
\end{remark}

Finally, in Theorem \ref{Th_Seq} we
generalise the results in Theorem 2 of Cavaliere \emph{et al.}\ (2018) by establishing the large sample properties of the adaptive IC-based estimator of the co-integration rank as in \eqref{IC(r)} using the lag length estimated by an 
information criterion as in \eqref{IC(k)} at the first step,
$\hat{r}_{\text{ALS-IC}}(\hat{k}_{\text{ALS-IC}})$.

\begin{theorem}
Let the conditions in Result 1 hold. Then it holds that $\hat{r}_{\mathrm{ALS}\text{-}\mathrm{IC}}(\hat{k}_{\mathrm{ALS}\text{-}\mathrm{IC}})\overset{p}{\to }r_0$, provided $c_T$ in
\eqref{IC(k)} and \eqref{IC(r)} satisfies the condition that $\frac{c_T}{T}+\frac{1}{c_T}
{\rightarrow} 0$ as $T \rightarrow \infty $.
\label{Th_Seq}
\end{theorem}

\begin{remark}
It is easy to show that the condition placed on $c_T$ in Theorem
\ref{Th_Seq} is not required if our purpose is to consistently
estimate the co-integration rank. Indeed, as was shown in Lemma
\ref{Th_joint_r}, any fixed lag length $k$ will also suffice in
that case. However, as shown  in Cavaliere \emph{et al.}\ (2018),
\emph{inter alia}, the finite sample performance of the
information criteria for determining the co-integration rank can
deteriorate badly if a fixed lag length is used which is not equal
to the true lag length, $ k_0$, and particularly so where it is
smaller than $ k_0$.
\end{remark}

\section{Numerical results}
\label{sec_MC}

In this section we use Monte Carlo simulation methods to
investigate the finite sample performance of the joint and
sequential adaptive methods for determining the co-integration
rank and autoregressive lag length outlined in Sections
\ref{sec_IC} and \ref{sec_Seq} and compare these with their
standard (non-adaptive) counterparts. The results from these Monte
Carlo experiments are reported in Tables 1-6. 

We will consider the following second-order VAR model of dimension
$p=2$ as our simulation DGP:
\begin{equation}
\Delta X_{t}=\alpha \beta^{\prime}X_{t-1}+\Gamma_{1}\Delta
X_{t-1}+\varepsilon_{t},   \ \ \
\alpha:=\left[
\begin{array}{cccc}
a & 0  \\
0 & b
\end{array}
\right],   \  \
\beta:=\left[
\begin{array}{cccc}
1 & 0  \\
0 & 1
\end{array}
\right],
\label{DGP}
\end{equation}
with $t=1-K,\ldots,T$,  $X_{-K}=\Delta X_{-K}=0$, where $K$ denotes the maximum lag order. 
In order to allow for true co-integration ranks, $ r_0$, of 0, 1 or
2, we set the parameters $a$ and $b$ in the long-run parameter
vector $\alpha$ in \eqref{DGP} as follows: $a=b=0$ for $r_0=0$,
$a=-0.4$ and $b=0$ for $r_0=1$, and $a=b=-0.4$ for $r_0=2$ (full
rank). Moreover, we set $\Gamma_{1}:=\gamma I_{2}$ with $\gamma
\in \{0, 0.1, 0.5, 0.9\}$.\footnote{For the simulation DGP in
\eqref{DGP}, it suffices that $ ( a, b, \gamma ) \in (-2,0]^2
\times [0,1)$ in order to satisfy the I($1,r$) conditions.}

We will consider three cases for the the innovation vector,
$\varepsilon_{t}$ in \eqref{DGP}.  The first case is that  $
\varepsilon_{t} \sim \mathrm{i.i.d.}\ N \left( 0, I_2\right) $ so that
$\varepsilon_{t}$ is homoskedastic.  This case will provide a
useful benchmark to investigate the effects of using adaptive
methods when they are not needed.
The second case considers conditionally heteroskedastic innovation processes, where the individual components of $\varepsilon_{t}$ follow the first-order AR stochastic
volatility [SV] model sets as $\varepsilon_{it}=v_{it}\exp {(h_{it})},$ $%
h_{it}=\lambda h_{it-1}+0.5\xi _{it}$, with $(\xi _{it},v_{it})^{\prime
}\sim \mathrm{i.i.d.}\ N(0,\mathrm{diag}(\sigma _{\xi }^{2},1))$, independent
across $i=1,2$. Results are reported for $\lambda =0.951$, $\sigma _{\xi
}=0.314$. This case constitutes a well-known conditionally heteroskedastic model for the innovations which has been used with the same parameter configuration in many other Monte Carlo experiments such as Gon\c{c}alves and
Kilian (2004), Cavaliere \emph{et al.}\ (2010), and Cavaliere \emph{et al.}\ (2015, 2018).
The third case we consider
sets $\varepsilon_{t}$ to be a non-stationary, unconditionally
heteroskedastic independent sequence of Gaussian variates,
characterised by a late positive variance shift. Specifically,
\begin{equation*}
\varepsilon_{t} \sim {N}\left( 0,\sigma
_{t}^{2}I_2\right), \ \text{ with } \sigma _{t}:= \left\{
\begin{array}{l l}
1 & \text{for } t\leq \left \lfloor 2T/3 \right \rfloor \\
3 & \text{for } t>\left \lfloor 2T/3 \right \rfloor
\end{array}
\right.\text{.}
\end{equation*}

In order to evaluate the behaviour of the adaptive and corresponding standard procedures in
practically relevant sample sizes we report results for 
$T=50$ and 100. All experiments are run over 1,000 Monte Carlo
replications and were programmed using MATLAB.
Our experiments are based on the no deterministic component case.
In all of our simulation experiments we set $K=4$ as the maximum
lag length considered.  Results for the joint information-based
estimates of the co-integration rank and lag length from Section
\ref{sec_IC} are reported first in Table 1, while results relating
to the sequential procedures from Section \ref{sec_Seq} are
reported in Tables 2 and 3 for the IC-based approaches in the case of SV innovations and single volatility break, respectively, and in Tables 4 and 5 for the sequential bootstrap-based procedures, again for the SV and single volatility break cases, respectively.
Finally, for comparison purposes, Table 6 reports the results for the joint information-based approaches in the homoskedastic case.

\medskip
\begin{center}
{\bf INSERT TABLES 1-6 HERE}
\end{center}
\medskip

Consider first Table 1 which reports results for determining the
co-integration rank $r$ (left two panels of Table 1) and the lag
order $k$ (right two panels of Table 1) using the joint
ALS-IC-based procedures detailed in Section \ref{sec_IC} together
with their corresponding standard information criteria-based
counterparts.  In particular, Table 1 reports the empirical
frequencies with which $ \tilde{r} $ and $ \tilde{k} $ from the
joint information-based estimator defined in \eqref{hat} select
the values $ r=0,1,2$ and $ k=1,2, 3, 4 $, respectively, for
each of the adaptive criteria ALS-HQC and ALS-BIC, and the
corresponding standard criteria, HQC and BIC.\footnote{We do not
consider the ALS-AIC estimator nor its standard counterpart in the
Monte Carlo experiments because the poor performance of AIC-based
approaches in finite samples is documented in many contributions
in the literature (see e.g., Kapetanios, 2004; Wang and Bessler,
2005; Cavaliere \emph{et al.}, 2015; Cavaliere \emph{et al.},
2016). Additional simulations show that, also in the case of
adaptive estimation, this criterion tends to  overestimate both
the true co-integration rank and the lag length. Nevertheless, the
adaptation with respect to the variance matrix profile
considerably improves the finite sample performance of the
AIC-based approach. These results are available on request.}

A number of observations can be made from the results reported in
Table 1.  Consider first the estimators of the co-integration
rank.

\begin{enumerate}

\item[(i)] In the case of autoregressive SV innovations reported
in the upper portion of Table 1, the performance of the adaptive
version of the information criteria, i.e.~ALS-HQC($k,p$) and
ALS-BIC($k,p$), is overall superior than (or at least as good as)
their standard counterparts, HQC($k,p$) and BIC($k,p$). The only
exception seen is for the case of no co-integration, $r_0=0$,
where the standard BIC outperforms its adaptive counterpart.
However, this is likely to be an artefact of the tendency of the
standard BIC to under-fit the `true' value of the co-integration
rank, which can be seen from the results in Table 1 for BIC and
ALS-BIC when $r_0>0$.

\item[(ii)] In the case of a single volatility break (lower
portion of Table 1), for a given penalty choice, i.e.~HQC or BIC,
the adaptive estimator is more efficacious, and often considerably
so, than the standard estimator in all but two of the cases
reported in Table 1.  As an example, while ALS-HQC selects the
correct value of $ r $ 81.4\% of the time when $r_0=0 $, $ \gamma
=0.5 $ and $T=100$, the standard HQC picks the correct rank only $
64.5\%$ of the time.

\item[(iii)] In the no co-integration case, $ r_0=0 $, the ALS-BIC
penalty delivers superior performance to the ALS-HQC, with the
same ordering holding for the approaches based on the standard BIC
and HQC criteria.  In particular we see that for both the adaptive
and standard cases the HQC penalty over-fits the co-integration
rank considerably more often than the BIC penalty. The degree of
over-fitting is, however, smaller for the ALS-HQC {\it
vis-\`a-vis} the standard HQC criterion.

\item[(iv)] For the case where $ r_0=1 $ there is overall little
to choose between the estimators based on the BIC and HQC
penalties; in particular, those based on the HQC penalty again
tend to over-fit the rank to a greater degree than those based on
the BIC penalty, with this effect again lessened for the adaptive
version of the estimator. In contrast, the BIC-based estimators
can tend to under-fit for $T=50$, excepting $ \gamma =0.9$.

\item[(v)] For the full rank case, $ r_0 = 2 $, the HQC-based
estimators generally select $ r_0 =2 $ more often than the
corresponding BIC-based estimators, although this is likely to
some degree to be an artefact of the tendency of the former to
over-fit, discussed above.

\item[(vi)] For all of the estimators considered, the lag length
and the magnitude of the lag parameter, $ \gamma$, can have a
considerable impact on the finite sample behaviour of the
co-integration rank estimators.  This impact appears to be less
pronounced, other things equal, for the adaptive variants of the
estimates and for the BIC-based procedures relative to the
corresponding HQC-based procedures.

\end{enumerate}

The following observations can also be made concerning the
behaviour of the estimators of the autoregressive lag length seen
in Table 1.

\begin{itemize}

\item[(i)] As observed above for co-integration rank estimation,
the adaptive information criteria outperform their standard
counterparts in selecting the true autoregressive lag length,
$k_0$, in almost all of the cases reported in Table 1. These
differences can again be large and generally tend to be larger,
other things equal, for the HQC penalty than for the BIC penalty.
As an example, while the ALS-HQC estimate of $k$ selects the
correct lag length 91.5\% of the time when $ r_0= 1 $, $ \gamma =
0.0$ and $T=100$, the standard HQC estimate selects the correct
lag length 71.7\% of the time.

\item[(ii)] The behaviour of each of the lag length estimators
considered is very similar, other things equal, across the three
values of the co-integration rank considered. Consequently, the
value of the true co-integration rank would appear to have
relatively little impact on the finite sample properties of the
lag length estimators.

\item[(iii)] The HQC-based procedures are superior to the
BIC-based procedures for $ \gamma = 0.1 $, presumably because of
the greater tendency of the HQC-based procedures to over-fit, a
tendency which is clearly seen for the larger values of $ \gamma $
considered, most notably with the non-adaptive versions of the
estimators.

\end{itemize}

\medskip

Let us now turn our attention to a discussion of the results in
Tables 2-5 which relate to the
sequential estimates from Section \ref{sec_Seq}. 

We first focus attention on the results reported for the two-step
IC-based procedures in Tables 2 and 3 for the cases of SV
innovations and a single volatility break, respectively. In
particular, we report the empirical frequencies with which both
standard and adaptive IC-based procedures select the lag length,
$k$, at the first step (`Step I' in the tables) and those with
which they select a co-integration rank, $r$, of zero, one or two
at the second step (`Step II' in the tables), using the lag length
estimated at the first step by each standard and adaptive
information criterion, IC($k,p$) and ALS-IC($k,p$).

%
The results for where the co-integration rank is determined using
the same information criterion at both steps of the sequential
procedure are overall similar to the results for the corresponding
joint IC-based approaches discussed above.  As an example, the
joint ALS-BIC estimate of $r$  in Table 1 selects the correct
co-integration rank 77.6\% (91.3\%) of the time when $r_0=1$,
$\gamma=0.5$ and $T=50$ ($T=100$) in the SV case, while the
corresponding sequential procedure based on ALS-BIC estimate at
both steps, i.e.\ ALS-BIC($\hat{k}_{\text{ALS-BIC}},r$), selects
the true rank 77.1\% (90.9\%) of the time.  Moreover, all of the
approaches considered appear to be fairly robust to the choice of
whether to use an adaptive or standard information criterion in
the first step of the sequential procedure as the results for the
co-integration rank determination appear very similar using either
ALS-IC($k,p$) or IC($k,p$).

%
%
%

\medskip

We now turn to a discussion of the results for the wild bootstrap
PLR procedure [denoted PLR-WB], together with the adaptive PLR
procedures of Boswijk and Zu (2022) implemented with either a
variance bootstrap [denoted ALR-VB] or a wild bootstrap [denoted
ALR-WB] in Tables 4 and 5 for the cases of SV innovations and a
single break in volatility, respectively. For each of these we
report the empirical frequencies with which they select a
co-integration rank, $r$, of zero, one or two.  We report results
for three case for the lag length used in these procedures.  The
first is an infeasible version based on knowledge of the true lag
length, i.e.\ we set $k=k_0$.  The other two select the lag length
in the first step of the two-step sequential procedure using
either standard BIC, $k=\hat{k}_{\text{BIC}}$, or its adaptive
counterpart, $k=\hat{k}_{\text{ALS-BIC}}$.\footnote{In Tables 4
and 5 we focus on BIC-based approaches for the selection of $k$
because these provide the best overall performance, see e.g.\
Cavaliere \emph{et al.}\ (2018). Moreover, we only report the
results for the lag length determination for the case of $r_0=1$.
The results for $r_0=0$ and 2 are very similar and thus, in the
interest of space, are not reported.}

A number of observations can be made from the results reported in
Tables 4 and 5.

\begin{itemize}

\item[(i)] In the no co-integration case, $ r_0 = 0 $, the PLR-WB
procedure is seen to have better ``size'' properties than either
of the ALR-VB and ALR-WB procedures which both tend to
over-estimate the co-integration rank to a greater degree than
does the PLR-WB procedure.  The behaviour of the ALR-VB or ALR-WB
procedures for $ r_0 =0 $ are fairly similar.

\item[(ii)] In the co-integrated case, $ r_0=1 $, the ALR-VB and
ALR-WB procedures both show a significantly higher empirical
probability of selecting the correct rank, $ r_0 =1 $, than does
the PLR-WB procedure which displays a tendency to under-fit the
co-integration rank, most notably for $ T=50$.  In the $ r_0=1 $
case the ALR-VB procedure appears to be slightly more efficacious
than the ALR-WB procedure.

\item[(iii)] In the full rank case, $ r_0 = 2 $, and with SV
innovations (Table 4), the adaptive procedures overall provide
slightly better performance than the PLR-WB procedure. Conversely,
in the single volatility break case (Table 5), the PLR-WB
procedure is, as in the zero rank case, more efficacious than
either the ALR-VB or ALR-WB procedures, both of which display a
consistent tendency to under-fit the rank.  As with the the $
r_0=1 $ case, the ALR-VB procedure appears to be slightly superior
to the ALR-WB procedure for both the heteroskedastic cases
considered.

\item[(iv)]  All of the PLR-WB, ALR-VB and ALR-WB procedures
appear to be fairly robust to the choice of the lag length made at
the first step of the sequential procedure. In particular, a
comparison of the results for $k=\hat{k}_{\text{BIC}}$ and
$k=\hat{k}_{\text{ALS-BIC}}$ with those for the corresponding
infeasible procedures based on a known lag length, $k=k_0$,
reveals that the loss in efficacy shown by the procedures for
determining the true co-integration rank at the second step due to
the estimation of the unknown lag length at the first step appears
very small and in some cases even negligible.  This is a
comforting result as it suggests there are only small losses in
finite sample efficacy from estimating the autoregressive lag
length, relative to an infeasible benchmark based on knowledge of
the true lag length.

\item[(v)] Focusing on the results for the lag length
determination reported in the right panel of Table 2, we can
observe that, overall, ALS-BIC appears to be more reliable than
the corresponding standard BIC. For example, in the case of a
single volatility break and $\gamma=0.5$ the selection frequency
of the true lag order, $\hat{k}=k_0=2$, for ALS-BIC is 82.0\%
(98.4\%) against 74.8\% (93.3\%) for BIC when $T=50$ ($T=100$).

\end{itemize}

\medskip

Finally, we investigate the potential losses of efficacy seen when
using the adaptive methods in the benchmark case of homoskedastic
innovations by comparing the results reported in Table 6 for the
adaptive IC-based methods with those of their standard
counterparts.  These results suggest that the performance of the
joint adaptive IC-based procedures do not deteriorate to any
significant degree when the shocks are homoskedastic, such that
the use of adaptive methods is unnecessary.  Indeed, when either
$r_0=1$ or 2, the performance of the ALS-IC-based approaches is
similar and sometimes even better than the results for their
corresponding standard counterparts. Conversely, in the case of no
co-integration, $r_0=0$, standard BIC and HQC-based approaches
outperform their adaptive counterparts. However, as pointed out
above, this is mainly an artefact of the overall tendency of the
standard criteria, especially BIC, to under-fit the true
co-integration rank.

\smallskip

To conclude this section, we compare the finite sample behaviour
of the adaptive information criteria-based methods with that of
the adaptive PLR test-based approaches. By comparing the results
reported in Tables 1, 2 and 3 with those in Tables 4 and 5, we
observe that, for the co-integrated case ($r_0=1$), the finite
sample performance of either joint or sequential ALS-IC is similar
to that of adaptive PLR test-based procedures. Conversely, when
$r_0=0$ the PLR test-based procedures outperform the adaptive
information criteria-based approaches, while this behaviour is
reversed when $r_0=2$ and $T=50$. In the case of full rank and
$T=100$, the performance of the methods considered are similar.
Finally, by comparing the results in Tables 1, 2 and 3 for the
joint and the sequential information criteria-based approaches for
selecting the lag length, we note that the ability of these
methods to determine $k$ are very similar.

\section{An Empirical Application: US Term Structure of Interest Rates}
\label{sec_emp}

In this section we provide an empirical application of the
adaptive information criteria-based approaches to the term
structure of interest rates in the US. In particular, we analyse
the time series $X_t = (X_{1t}, \ldots, X_{5t})^\prime$ of monthly
zero yields from January 1970 to December 2012, for maturities
equal to 3 months ($X_{1t}$), 1 year ($X_{2t}$), 3 years
($X_{3t}$), 5 years ($X_{4t}$), and 10 years ($X_{5t}$).

The co-integration analysis of $X_t$ has already been considered
by Boswijk \emph{et al.}\ (2016) and Boswijk and Zu (2022).
In particular, in order to account for the unconditional heteroskedasticity
present in the data,  sequential procedures where the lag length is selected at the first step according to (standard) HQC($k,p$), 
and then the co-integration rank of the system is determined using either PLR-WB 
 (Boswijk \emph{et al.}, 2016) or adaptive PLR tests 
(Boswijk and Zu, 2022) were adopted.
Here we apply the adaptive information-based methods to estimate
the co-integration rank and autoregressive lag order of the system
and compare these results with those obtained in the two previous
analyses cited above. In what follows, the VAR models are fitted
with a restricted trend and, for all methods, the maximum number
of lags considered is $K=4$. The number of bootstrap samples used
in the bootstrap algorithms is $B=999$.

We first focus on the joint determination of the co-integration
rank and lag length using adaptive joint information
criterion-based procedures as outlined in Section \eqref{sec_IC}
and the standard counterparts. These results are reported in Table
7. The results in Table 7 show that all of the joint information
criteria, both adaptive and non-adaptive, agree on selecting a lag
length of $\tilde{k}=2$. Moreover, both standard and adaptive
versions of the joint BIC-based approach delivers the same
estimate of the co-integration rank, namely
$\tilde{r}_{\text{BIC}}=\tilde{r}_{\text{ALS-BIC}}=2$. Conversely,
the joint HQC-based approaches select a higher co-integration
rank. Specifically, the  co-integration rank selected using the
(standard) joint HQC-based approach is 3, i.e.,
$\tilde{r}_{\text{HQC}}= 3$, whereas
$\tilde{r}_{\text{ALS-HQC}}=4$ is obtained using the adaptive
version.

\medskip
\begin{center}
{\bf INSERT TABLE 7 HERE}
\end{center}
\medskip

We now consider in Table 8 the results obtained using the
sequential procedures for determining the lag length and then the
co-integration rank. In particular, the upper panel of Table 8
shows the results for the selection of $k$ in the first step of
the sequential procedure, whereas the results for the
determination of $r$ at the second step using information criteria
and PLR tests are reported in the middle and lower panels of Table
8, respectively. Note that the results reported in the lower panel
of Table 8 for the case of $\hat{k}=2$ reproduce those in Boswijk
\emph{et al.}\ (2016) and Boswijk and Zu (2022) who use standard
HQC to select the lag length and therefore they set $\hat{k}=2$.
The results for the first step of the sequential procedure show
that all but the standard BIC information criteria agree on a
choice of $\hat{k}=2$; standard BIC chooses
$\hat{k}_{\text{BIC}}=1$. Therefore, on balance, we would
recommend a VAR model of order 2.

Let us next focus on the second step of the sequential procedure
and, in particular, on the determination of the co-integration
rank obtained by the PLR tests (see the lower panel of Table 8).
For $\hat{k}=2$, the results for the adaptive and non-adaptive
bootstrap-based PLR test procedures vary according to the nominal
significance level considered. In particular, at a standard 5\%
level we select $\hat{r}=2$ using the (non-adaptive) PLR-WB
procedure, whereas the two adaptive PLR methods yield $\hat{r}=4$,
again replicating the results in Boswijk \emph{et al.}\ (2016) and
Boswijk and Zu (2022), respectively. Using a 1\% significance
level, we still select a co-integration rank of 2 using the
(standard) PLR-WB but we would now select $\hat{r}=3$ using the
two adaptive PLR test-based procedures. The results for the
information criteria used in the second step of the sequential
procedure show that, using $\hat{k}=2$, HQC-based approaches in
both adaptive and non-adaptive form agree with the selection of
$\hat{r}=4$ also made at the 5\% level made by the adaptive PLR
test-based procedures.  The co-integration rank of $\hat{r}=2$
selected using both the adaptive and non-adaptive BIC-based
approaches matches that chosen by the (non-adaptive) PLR-WB test
procedure. It is worth noting that, when setting $\hat{k}=1$ as
suggested by the (standard) BIC($k,p$), the results for both
information criteria and PLR tests in step 2 of the sequential
procedure are much more variable across the methods with the rank
selected anywhere between 2 and 5. Therefore, we would not
recommend the conclusions based on $\hat{k}=1$. In particular,
because  BIC uses a stricter penalty term than HQC, we would
expect, other things equal, that BIC-based approaches will often
select a lower lag length and/or co-integration rank than
HQC-based approaches. Moreover, this tendency of standard BIC
might be exacerbated by the presence of heteroskedasticity in the
data, thus allowing the adaptation with the respect to the
volatility process to deliver more reliable results in small
samples.

\medskip
\begin{center}
{\bf INSERT TABLE 8 HERE}
\end{center}
\medskip

In summary, overall our results seem strongly in favour of a
selection of an autoregressive lag length of 2. However, the
selected co-integration rank varies according to the method used.
In particular, the joint and sequential (for $\hat{k}=2$)
HQC-based approaches select a co-integration rank of 4, while the
joint and sequential (for $\hat{k}=2$) BIC-based approaches select
rank 2. The sequential procedures based on PLR tests and
$\hat{k}=2$ select $\hat{r}=2$ in non-adaptive form, $\hat{r}=3$
when using a 1\% significance level and $\hat{r}=4$ when using a
5\% significance level. This is in some ways consistent with the
findings for BIC and HQC-based methods since decreasing the
significance level is qualitatively the same as using a stricter
penalty in the information criterion. Finally, it is worth noting
that the choice of a rank equal to 4 implies the presence of a
single stochastic trend driving the five yields and is in line
with the (weak-form) expectation hypothesis of interest rates
(see, for example, Campbell and Shiller, 1987), which implies that
the (long-term) level factor - but not the slope nor the curvature
- of the interest rate yield curve is a random walk process, so
that $\beta^\prime X_t$ consists of spreads $X_{it}-X_{1t}$ for $i
=2, 3, 4, 5$.

\section{Conclusions}
\label{sec_conc}

In this paper we have proposed new methods for determining the
co-integration rank and the lag order in heteroskedastic VAR
models which exploit the time variation in the unconditional error
variance matrix. In particular, we have proposed adaptive
information criteria-based approaches to jointly determine the
co-integration rank and the autoregressive lag length. Provided
standard conditions hold on the penalty term hold, these methods
are proved to be weakly consistent for co-integration rank and lag
order determination.  We have also demonstrated that the adaptive
PLR rank determination procedure of Boswijk and Zu (2022),
originally developed under the assumption of a known
autoregressive lag length, remains asymptotically valid when a
consistent lag length estimate, such as that provided by an
adaptive information criterion, is used.  Monte Carlo experiments
reported indicate that the adaptive information criteria-based
approaches generally outperform standard methods in finite samples
when non-stationary volatility is present in the data.

\bigskip

\section*{Acknowledgments}
This paper is dedicated to the memory of our dear friend and colleague Mike McAleer.
The authors wish to thank Anders Rahbek and Yang Zu  for their helpful
comments and suggestions. 
The authors also thank participants at the Computation and Financial Econometrics in London (December 2015 and 2017), the Bootstrap Workshop at the Amsterdam School of Economics (November 2015), and the RCEA Time Series Econometrics Workshop in Rimini (June 2016).
This research was supported by the Danish Council for Independent Research (DSF Grant 015-00028B) and the Italian Ministry of University and Research (PRIN 2017
Grant 2017TA7TYC).

\bigskip

\section*{References}

\setcounter{section}{0}

\begin{description}
\item {Akaike, H.\ }(1974): A new look at the statistical model
identification, \emph{IEEE Transactions on Automatic Control} 19, 716--723.

\item {Asai, M., M.\ McAleer and J.\ Yu }(2006): Multivariate Stochastic Volatility: A Review, \emph{Econometric Reviews} 25, 145--175.


\item {Boswijk, H.\ P.\ }(1995):
Identifiability of cointegrated systems, Tinbergen Institute Discussion
Paper \# 95-78, \url{http://dare.uva.nl/document/2/163397}.

\item {Boswijk, H.\ P., G.\ Cavaliere,  A.\ Rahbek and A.\ M.\ R.\ Taylor }(2016):
Inference on co-integration parameters in heteroskedastic vector
autoregressions,  \emph{Journal of Econometrics} 192, 64--85.

\item Boswijk, H.\ P.\ and P.\ H.\ Franses (1992): Dynamic
specification and cointegration, \emph{Oxford Bulletin of
Economics and Statistics} 54, 369--381.



\item Boswijk, H.\ P.\ and Y.\ Zu (2022): Adaptive testing for cointegration with nonstationary volatility, \emph{Journal of Business \& Economic Statistics}, forthcoming;
\url{https://doi.org/10.1080/07350015.2020.1867558}.


\item Campbell, J.\ Y.\ and R.\ J.\ Shiller (1987): Cointegration and tests of present value model, \emph{Journal of Political Economy} 95, 1062--1088.



\item {Cavaliere, G. and A.\ M.\ R.\ Taylor }(2008): Bootstrap unit root tests for time series with nonstationary volatility, \emph{Econometric Theory} 24, 43--71.

\item {Cavaliere, G., L.\ De Angelis, A.\ Rahbek and A.\ M.\ R.\ Taylor }%
(2018): Determining the co-integration rank in heteroskedastic VAR
models of unknown order, \emph{Econometric Theory}, 34, 349--382.

\item {Cavaliere, G., L.\ De Angelis, A.\ Rahbek and A.\ M.\ R.\ Taylor }%
(2015): A comparison of sequential and information-based methods
for determining the co-integration rank in heteroskedastic VAR
models, \emph{Oxford Bulletin of Economics and Statistics} 77,
106--128.


\item {Cavaliere, G., A.\ Rahbek and A.\ M.\ R.\ Taylor } (2010):
Testing for co-integration in vector autoregressions with non-stationary
volatility, \emph{Journal of Econometrics} 158, 7--24.

\item {Cavaliere, G., A.\ Rahbek and A.\ M.\ R.\ Taylor }(2012):
Bootstrap determination of the co-integration rank in VAR models,
\emph{Econometrica } 80, 1721--1740.

\item Cavaliere, G., A.\ Rahbek and  A.\ M.\ R.\ Taylor (2014)
Bootstrap determination of the co-integration rank in
heteroskedastic VAR models. \emph{Econometric Reviews} 33,
606--650.


%
%
%
\item {Cheung, Y.\ W.\ and K.\ S.\ Lai }(1993): Finite-sample sizes of Johansen's
likelihood ratio tests for cointegration, \emph{Oxford Bulletin of Economics
and Statistics} 55, 313--328.


\item Gon\c{c}alves, S.\ and L.\ Kilian (2004): Bootstrapping
autoregressions with conditional heteroskedasticity of unknown form, \emph{%
Journal of Econometrics }123, 89--120.

\item {Hannan, E.\ J.\ and B.\ G.\ Quinn }(1979): The determination of the order
of an autoregression, \emph{Journal of the Royal Statistical Society, Series
B} 41, 190--195.

\item Hansen, P.\ R.\ (2002): Generalized reduced rank regression, Brown University Working
Paper No.\ 2002-02, \url{http://ssrn.com/abstract=302859}.

\item Hansen, P.\ R.\ (2003): Structural changes in the cointegrated vector autoregressive
model, \emph{Journal of Econometrics} 114, 261--295.

\item {Haug, A.\ A.\ }(1996): Test for cointegration: A Monte Carlo comparison,
\emph{Journal of Econometrics} 71, 89--115.



\item {Johansen, S.\ }(1996): \emph{Likelihood-Based Inference in
Cointegrated Vector Autoregressive Models}, Oxford: Oxford
University Press.

%
%
\item {Kapetanios, G.\ }(2004): The asymptotic distribution of the
cointegration rank estimator under the Akaike information criterion, \emph{%
Econometric Theory} 20, 735--742.
%

%

\item {L\"{u}tkepohl, H.\ and P.\ Saikkonen }(1999): Order selection in
testing for the cointegrating rank of a VAR process, In: \emph{%
Cointegration, Causality, and Forecasting. A Festschrift in Honour
of Clive W.\ J.\ Granger}, Engle, R.\ F.\ and H.\ White (eds.), Oxford:
Oxford University Press, 168--199.

\item {McAleer, M. }(2005): Automated inference and learning in modeling financial volatility. \emph{Econometric Theory} 21(1), 232--261.

\item {McAleer, M., S.\ Hoti and F.\ Chan }(2009): Structure and Asymptotic Theory for Multivariate Asymmetric Conditional Volatility, \emph{Econometric Reviews} 28, 422--440.

\item {McAleer, M.\ and M.\ C.\ Medeiros }(2008): Realized Volatility: A Review, \emph{Econometric Reviews} 27, 10--45.

\item {McConnell, M.\ M.\ and G.\ Perez Quiros }(2000): Output fluctuations in
the United States: what has changed since the early 1980s?, \emph{American
Economic Review} 90, 1464--1476.

\item {Nielsen, B.\ }(2006): Order determination in general vector
autoregressions, \emph{IMS Lecture Notes -- Monograph Series} 52,
93--112.

\item {Patilea, V.\ and H.\ Ra\"{i}ssi} (2012): Adaptive estimation of vector autoregressive models with time-varying variance: Application to testing linear causality in mean, \emph{Journal of Statistical Planning and Inference} 142, 2891--2912.


%

%





\item {Schwarz, G.\ }(1978): Estimating the dimension of a model, \emph{%
Annals of Statistics} 6, 461--464.

\item {Sensier, M.\ and D.\ van Dijk }(2004): Testing for volatility changes
in U.S. macroeconomic time series, \emph{Review of Economics and Statistics}
86, 833--839.

%
%

\item {Wang, Z.\ and D.\ A.\ Bessler }(2005): A Monte Carlo study on the
selection of cointegration rank using information criteria, \emph{%
Econometric Theory} 21, 593--620.

\item {Xu, K.-L.\ and P.\ C.\ B.\ Phillips} (2008): Adaptive estimation of autoregressive models with time-varying variances, \emph{Journal of Econometrics} 142, 265--280.

\end{description}

%
%
%

\pagebreak

\newpage

\appendix

\setcounter{equation}{0}\renewcommand{\theequation}{A.\arabic{equation}}%
\renewcommand{\thelemma}{A.\arabic{lemma}}\setcounter{lemma}{0} %

\section{Appendix}
\label{app}

\noindent \textbf{Notation and preliminary results.} Write the unrestricted
model (with $r=p$ and $k=K$) without deterministic terms as%
\begin{equation*}
\Delta X_{t}=\Pi X_{t-1}+\Psi Z_{t}+\varepsilon _{t}=[\Pi :\Psi
]W_{t}+\varepsilon _{t}=\left( W_{t}^{\prime }\otimes I_{p}\right) \theta
+\varepsilon _{t},
\end{equation*}%
where $W_{t}=(X_{t-1}^{\prime },Z_{t}^{\prime })^{\prime }$ with $%
Z_{t}=Z_{t}^{(K)}=(\Delta X_{t-1}^{\prime },\ldots ,\Delta X_{t-K+1}^{\prime
})^{\prime }$, and where $\theta =\limfunc{vec}[\Pi :\Psi ]$.

The lag order restriction $k<K$ implies particular zeros on $\limfunc{vec}%
\Psi $, say,%
\begin{equation*}
\limfunc{vec}\Psi =\left( 
\begin{array}{c}
\psi _{1}^{(k)} \\ 
0%
\end{array}%
\right) .
\end{equation*}%
The cointegration restriction $\limfunc{rank}\Pi \leq r$ implies $\Pi
=\alpha \beta ^{\prime }$, and hence%
\begin{equation*}
\limfunc{vec}\Pi =\limfunc{vec}(\alpha \beta ^{\prime })=(I_{p}\otimes
\alpha )\limfunc{vec}\left( \beta ^{\prime }\right) ,
\end{equation*}%
where $\alpha $ and $\beta $ are $p\times r$ matrices. Depending on $r$, we
normalise $\beta $ as $c^{\prime }\beta =I_{r}$, for some known $p\times r$
matrix $c$ of full column rank. Defining $c_{\perp }$ as the orthogonal
complement of $c$, and $\bar{c}=c(c^{\prime }c)^{-1}\,$, this leads to $%
\beta =\bar{c}+c_{\perp }\Phi ^{\prime }$ for some $r\times (p-r)$ matrix $%
\Phi $ of unknown parameters; hence%
\begin{equation}
\limfunc{vec}\Pi =(I_{p}\otimes \alpha )(\limfunc{vec}(\bar{c}^{\prime
})+(c_{\perp }\otimes I_{r})\phi )=:g^{(r)}(\phi ,\alpha ),  \label{g}
\end{equation}%
where $\phi =\limfunc{vec}\Phi $ and the function $g^{(r)}$ is implicitly
defined.

With known $\Sigma _{t}$, minus two times the log-likelihood of the
unrestricted model, up to an additive constant, is given by%
\begin{eqnarray*}
-2\ell _{T}(\theta ) &=&\sum_{t=1}^{T}\left( \Delta X_{t}-\left(
W_{t}^{\prime }\otimes I_{p}\right) \theta \right) ^{\prime }\Sigma
_{t}^{-1}\left( \Delta X_{t}-\left( W_{t}^{\prime }\otimes I_{p}\right)
\theta \right) \\
&=&\sum_{t=1}^{T}\left( \hat{\varepsilon}_{t}-\left( W_{t}^{\prime }\otimes
I_{p}\right) (\theta -\hat{\theta})\right) ^{\prime }\Sigma _{t}^{-1}\left( 
\hat{\varepsilon}_{t}-\left( W_{t}^{\prime }\otimes I_{p}\right) (\theta -%
\hat{\theta})\right) \\
&=&\sum_{t=1}^{T}\hat{\varepsilon}_{t}^{\prime }\Sigma _{t}^{-1}\hat{%
\varepsilon}_{t}+(\theta -\hat{\theta})^{\prime }\sum_{t=1}^{T}\left(
W_{t}W_{t}^{\prime }\otimes \Sigma _{t}^{-1}\right) (\theta -\hat{\theta}),
\end{eqnarray*}%
where $\hat{\theta}$ is the unrestricted MLE%
\begin{equation*}
\hat{\theta}=\left[ \sum_{t=1}^{T}\left( W_{t}W_{t}^{\prime }\otimes \Sigma
_{t}^{-1}\right) \right] ^{-1}\sum_{t=1}^{T}\left( W_{t}\otimes \Sigma
_{t}^{-1}\right) \Delta X_{t},
\end{equation*}%
and $\hat{\varepsilon}_{t}=\Delta X_{t}-\left( W_{t}^{\prime }\otimes
I_{p}\right) \hat{\theta}$. Estimating different submodels $(r,k)$ involves
minimizing $-2\ell _{T}(\theta )$ over $\theta $ under the restriction%
\begin{equation*}
\theta ^{(k,r)}=\left( 
\begin{array}{c}
g^{(r)}(\phi ,\alpha ) \\ 
\psi _{1}^{(k)} \\ 
0%
\end{array}%
\right) ,
\end{equation*}%
which yields the restricted estimator $\tilde{\theta}^{(k,r)}$.

Using the true value $\beta _{0}$ and hence $r_{0}$, define%
\begin{equation*}
D_{T}=\left[ 
\begin{array}{ccc}
T^{-1}c_{\perp } & T^{-1/2}\beta _{0} & 0 \\ 
0 & 0 & T^{-1/2}I_{p(K-1)}%
\end{array}%
\right] \otimes I_{p},
\end{equation*}%
such that%
\begin{equation*}
D_{T}^{\prime }\left( W_{t}\otimes I_{p}\right) =T^{-1/2}\left( 
\begin{array}{c}
T^{-1/2}c_{\perp }^{\prime }X_{t-1} \\ 
\beta _{0}^{\prime }X_{t-1} \\ 
Z_{t}%
\end{array}%
\right) \otimes I_{p}.
\end{equation*}%
This is used to normalise the factors of the log-likelihood ratio function: 
\begin{eqnarray*}
\Lambda _{T}(\theta ) &:=&-2\left[ \ell _{T}(\theta )-\ell _{T}(\hat{\theta})%
\right] \\
&=&(\theta -\hat{\theta})^{\prime }\sum_{t=1}^{T}\left( W_{t}W_{t}^{\prime
}\otimes \Sigma _{t}^{-1}\right) (\theta -\hat{\theta}) \\
&=&(\theta -\hat{\theta})^{\prime }D_{T}^{\prime -1}\left[
\sum_{t=1}^{T}D_{T}^{\prime }\left( W_{t}W_{t}^{\prime }\otimes \Sigma
_{t}^{-1}\right) D_{T}\right] D_{T}^{-1}(\theta -\hat{\theta}).
\end{eqnarray*}%
Note that $D_{T}$ has been defined such that $D_{T}^{-1}(\theta -\hat{\theta}%
)$ and the normalised observed information matrix in square brackets are
bounded in probability (and the latter has a non-singilar limit). Indeed, as
shown by Boswijk and Zu (2022), 
\begin{equation*}
D_{T}^{-1}(\hat{\theta}-\theta _{0})\overset{w}{\rightarrow }\left[ 
\begin{array}{cc}
J_{1} & 0 \\ 
0 & J_{2}%
\end{array}%
\right] ^{-1}\left( 
\begin{array}{c}
S_{1} \\ 
S_{2}%
\end{array}%
\right) ,
\end{equation*}%
and 
\begin{equation*}
\sum_{t=1}^{T}D_{T}^{\prime }\left( W_{t}W_{t}^{\prime }\otimes \Sigma
_{t}^{-1}\right) D_{T}\overset{w}{\rightarrow }\left[ 
\begin{array}{cc}
J_{1} & 0 \\ 
0 & J_{2}%
\end{array}%
\right] ,
\end{equation*}%
where $S_{1}$ and $J_{1}$ are the limits of the normalised score vector and
information matrix of the cointegration parameters $\phi $, and $S_{2}$ and $%
J_{2}$ are the corresponding limits for the remaining parameters ($\alpha $
and $\Psi $). Furthermore, Boswijk and Zu (2022) show that the same limit
results apply if the true sequence $\left\{ \Sigma _{t}\right\} _{t=1}^{T}$
is replaced by the non-parametric estimate $\{\hat{\Sigma}_{t}\}_{t=1}^{T}$
in the expression for $\ell _{T}$ and hence $\Lambda _{T}$.

Extending the above results to the case with a (possibly restricted)
constant or linear trend term requires $X_{t-1}$ and possibly $Z_{t}$ to be
extended by such deterministic terms, and a corresponding extension of the
matrix $D_{T}$. This will not be considered explicitly here.

Finally, it will be convenient to define $\mathrm{LR}(\mathcal{H}%
_{k_{1},r_{1}}|\mathcal{H}_{k_{2},r_{2}}):=\Lambda _{T}(\tilde{\theta}%
^{(k_{1,}r_{1})})-\Lambda _{T}(\tilde{\theta}^{(k_{2,}r_{2})})$, the
likelihood ratio statistic for $\mathcal{H}_{k_{1},r_{1}}$ against $\mathcal{%
H}_{k_{2},r_{2}}$, where $(k_{1},r_{1})$ and $(k_{2},r_{2})$ are particular
values of $(k,r)$ with $k_{1}\leq k_{2}$ and $r_{1}\leq r_{2}$. \hfill $%
\square $

\bigskip

\noindent \textbf{Proof of Lemma 1.} To obtain the results of Lemma 1, we
proceed to analyse 
\begin{equation*}
\mathrm{ALS}\text{-}\mathrm{IC}(k,r)-\mathrm{ALS}\text{-}\mathrm{IC}%
(k,r_{0})=\Lambda _{T}(\tilde{\theta}^{(k,r)})-\Lambda _{T}(\tilde{\theta}%
^{(k,r_{0})})+c_{T}\left[ \pi (k,r)-\pi (k,r_{0})\right] ,
\end{equation*}%
where $\pi (k,r)=r(2p-r)+p^{2}(k-1)$. We first consider the case where $%
k\geq k_{0}$, such that the chosen lag length is well- (or over-) specified.
After that, we consider the case of under-specified dynamics ($k<k_{0}$).

When $k\geq k_{0}$, then $\mathcal{H}_{k,r_{0}}$ is a well-specified model,
and hence $\Lambda _{T}(\tilde{\theta}^{(k,r_{0})})$ is the LR\ statistic
for the null hypothesis that the lag length is (less than or) equal to $k$
and the cointegrating rank is $r_{0}$ in the unrestricted model. As this
null hypothesis is true, $\Lambda _{T}(\tilde{\theta}^{(k,r_{0})})$ will
have a limiting null distribution, being the distribution of the sum of the
LR statistic in Boswijk and Zu (2022) and a $\chi ^{2}$ random variable.
Most importantly, $\Lambda _{T}(\tilde{\theta}^{(k,r_{0})})=O_{p}(1)$.

For $r>r_{0}$, we have $\Lambda _{T}(\tilde{\theta}^{(k,r)})-\Lambda _{T}(%
\tilde{\theta}^{(k,r_{0})})=-\mathrm{LR}(\mathcal{H}_{k,r_{0}}|\mathcal{H}%
_{k,r})$, which is minus the LR statistic for a true null hypothesis in an
overspecified model, and hence it is $O_{p}(1)$. Because $\pi (k,r)-\pi
(k,r_{0})>0$, it follows that 
\begin{equation*}
\Pr \left( \mathrm{ALS}\text{-}\mathrm{IC}(k,r)-\mathrm{ALS}\text{-}\mathrm{%
IC}(k,r_{0})>0\right) \rightarrow 1,
\end{equation*}%
provided $c_{T}\rightarrow \infty $.

For $r<r_{0}$, we have $\Lambda _{T}(\tilde{\theta}^{(k,r)})-\Lambda _{T}(%
\tilde{\theta}^{(k,r_{0})})=\mathrm{LR}(\mathcal{H}_{k,r}|\mathcal{H}%
_{k,r_{0}})$. In this case, the null hypothesis is violated, which will
cause the statistic to diverge (to $+\infty $) at the rate $O_{p}(T)$. To
obtain this rate, consider first the simplest case where $r=0$ and $%
k_{0}=k=K=1$, so that the estimator of $\Psi $ is zero under both
constraints, and $\tilde{\theta}^{(k,r)}=\limfunc{vec}\tilde{\Pi}^{(k,r)}=0$%
. Therefore, 
\begin{eqnarray*}
\mathrm{LR}(\mathcal{H}_{k,r}|\mathcal{H}_{k,r_{0}}) &=&\Lambda _{T}(\tilde{%
\theta}^{(k,r)})-\Lambda _{T}(\tilde{\theta}^{(k,r_{0})}) \\
&=&\hat{\theta}^{\prime }D_{T}^{\prime -1}\left[ \sum_{t=1}^{T}D_{T}^{\prime
}\left( W_{t}W_{t}^{\prime }\otimes \Sigma _{t}^{-1}\right) D_{T}\right]
D_{T}^{-1}\hat{\theta}-\Lambda _{T}(\tilde{\theta}^{(k,r_{0})}).
\end{eqnarray*}%
Since $D_{T}^{-1}\hat{\theta}=D_{T}^{-1}\theta _{0}+O_{p}(1)$, with%
\begin{equation*}
D_{T}^{-1}\theta _{0}=\left( \left[ 
\begin{array}{c}
T(\beta _{0\perp }^{\prime }c_{\perp })^{-1}\beta _{0\perp }^{\prime } \\ 
T^{1/2}(c^{\prime }\beta _{0})^{-1}c^{\prime }%
\end{array}%
\right] \otimes I_{p}\right) \left[ \beta _{0}\otimes I_{p}\right] \limfunc{%
vec}\alpha _{0}=\left( 
\begin{array}{c}
0 \\ 
T^{1/2}\limfunc{vec}\alpha _{0}%
\end{array}%
\right) ,
\end{equation*}%
and $\Lambda _{T}(\tilde{\theta}^{(k,r_{0})})$ is $O_{p}(1)$ as before, this
leads to $\Lambda _{T}(\tilde{\theta}^{(k,r)})-\Lambda _{T}(\tilde{\theta}%
^{(k,r_{0})})=O_{p}(T)$. More generally, we will find that the divergence
rate of 
\begin{equation*}
\Lambda _{T}(\tilde{\theta}^{(k,r)})=(\tilde{\theta}^{(k,r)}-\hat{\theta}%
)^{\prime }D_{T}^{\prime -1}\left[ \sum_{t=1}^{T}D_{T}^{\prime }\left(
W_{t}W_{t}^{\prime }\otimes \Sigma _{t}^{-1}\right) D_{T}\right] D_{T}^{-1}(%
\tilde{\theta}^{(k,r)}-\hat{\theta})
\end{equation*}%
will be determined by 
\begin{equation*}
D_{T}^{-1}(\tilde{\theta}^{(k,r)}-\hat{\theta})=D_{T}^{-1}(\tilde{\theta}%
^{(k,r)}-\theta _{0})+O_{p}(1),
\end{equation*}%
and since $\theta _{0}$ does not lie in the constrained parameter space such
that the difference $\tilde{\theta}^{(k,r)}-\theta _{0}$ will be $O_{p}(1)$
but not $o_{p}(1)$, it follows that $D_{T}^{-1}(\tilde{\theta}^{(k,r)}-\hat{%
\theta})=O_{p}(T^{1/2})$ as before, and hence $\Lambda _{T}(\tilde{\theta}%
^{(k,r)})=O_{p}(T)$. The term $c_{T}\left[ \pi (k,r)-\pi (k,r_{0})\right] $
is negative and diverges at the rate $c_{T}$; therefore 
\begin{equation*}
\Pr \left( \mathrm{ALS}\text{-}\mathrm{IC}(k,r)-\mathrm{ALS}\text{-}\mathrm{%
IC}(k,r_{0})>0\right) \rightarrow 1
\end{equation*}%
provided $c_{T}/T\rightarrow 0$.

Next, consider the case $k<k_{0}$, so that we are comparing two
(dynamically) misspecified models. When $r>r_{0}$, such that the larger
model encompasses the correct cointegration rank, we may use the following
decomposition:%
\begin{equation}
\mathrm{LR}(\mathcal{H}_{k,r_{0}}|\mathcal{H}_{k,r})=\mathrm{LR}(\mathcal{H}%
_{K,r_{0}}|\mathcal{H}_{K,r})+\mathrm{LR}(\mathcal{H}_{k,r_{0}}|\mathcal{H}%
_{K,r_{0}})-\mathrm{LR}(\mathcal{H}_{k,r}|\mathcal{H}_{K,r}),  \label{decomp}
\end{equation}%
which follows from $\mathcal{H}_{k,r_{0}}\subset \mathcal{H}%
_{K,r_{0}}\subset \mathcal{H}_{K,r}$ and $\mathcal{H}_{k,r_{0}}\subset 
\mathcal{H}_{k,r}\subset \mathcal{H}_{K,r}$, and equating the sum of the LR
statistics for both nested sequences of hypotheses. The first right-hand
side expression in (\ref{decomp}) is the LR statistic for the correct
cointegration rank in a well-specified model, and hence $O_{p}(1)$. The
second and third terms in (\ref{decomp}) are LR statistics for an incorrect
lag length against an unrestricted lag length. Both test statistics will
diverge, but their difference is $O_{p}(1)$, as we will now show.

Recall the definition of $g^{(r)}(\phi ,\alpha )$ in (\ref{g}), and define
the corresponding Jacobian matrix%
\begin{equation*}
G^{(r)}(\phi ,\alpha )=\left[ \frac{\partial g^{(r)}(\phi ,\alpha )}{%
\partial \phi ^{\prime }}:\frac{\partial g^{(r)}(\phi ,\alpha )}{\partial 
\limfunc{vec}(\alpha )^{\prime }}\right] =\left[ (c_{\perp }\otimes \alpha
):\left( \beta \otimes I_{p}\right) \right] ,
\end{equation*}%
where $\beta $ is determined from $\phi $ as $\limfunc{vec}\beta ^{\prime }=%
\limfunc{vec}((c^{\prime }c)^{-1}c^{\prime })+(c_{\perp }\otimes I_{r})\phi $%
. Next, define the Jacobian matrices evaluated at the true values%
\begin{equation*}
G_{0}=G^{(r_{0})}(\phi _{0},\alpha _{0}),\qquad G=G^{(r)}(\phi
_{0}^{(r)},\alpha _{0}^{(r)}).
\end{equation*}%
Here $(\phi _{0},\alpha _{0})$ is the true parameter value in the model $%
\mathcal{H}_{K,r_{0}}$, and similarly $(\phi _{0}^{(r)},\alpha _{0}^{(r)})$
is the true value in the overspecified model $\mathcal{H}_{K,r}$ with $%
r>r_{0}$. Note that $\phi $ and $\alpha $ are not identified in the
over-specified model, but one can choose a true value such that $\limfunc{vec%
}\Pi _{0}=g^{(r)}(\phi _{0}^{(r)},\alpha _{0}^{(r)})$. Using a linearisation
of the rank-restricted model, and hence a quadratic approximation of the
log-likelihood, we have%
\begin{equation*}
\tilde{\theta}^{(K,r_{0})}-\theta _{0}=\left( 
\begin{array}{c}
\tilde{\pi}^{(K,r_{0})}-\pi _{0} \\ 
\tilde{\psi}^{(K,r_{0})}-\psi _{0}%
\end{array}%
\right) =\left[ 
\begin{array}{cc}
G_{0} & 0 \\ 
0 & I%
\end{array}%
\right] \left( \sum_{t=1}^{T}\mathbb{W}_{t}^{0}\mathbb{W}_{t}^{0\prime
}\right) ^{-1}\sum_{t=1}^{T}\mathbb{W}_{t}^{0}z_{t}+o_{p}(T^{-1/2}),
\end{equation*}%
where $z_{t}=\sigma _{t}^{-1}\varepsilon _{t}$ (with $\sigma _{t}$ the
symmetric square root of $\Sigma _{t}$) and 
\begin{equation*}
\mathbb{W}_{t}^{0}=\left( 
\begin{array}{c}
\mathbb{W}_{0t}^{0} \\ 
\mathbb{W}_{1t} \\ 
\mathbb{W}_{2t}%
\end{array}%
\right) =\left( 
\begin{array}{c}
G_{0}^{\prime }(X_{t-1}\otimes \sigma _{t}^{-1}) \\ 
Z_{1t}\otimes \sigma _{t}^{-1} \\ 
Z_{2t}\otimes \sigma _{t}^{-1}%
\end{array}%
\right) .
\end{equation*}%
Here the vector of lagged differences $Z_{t}$ has been partitioned into the
retained lags $Z_{1t}=(\Delta X_{t-1}^{\prime },\ldots ,\Delta
X_{t-k+1}^{\prime })^{\prime }$ and the excluded lags $Z_{2t}=(\Delta
X_{t-k}^{\prime },\ldots ,\Delta X_{t-K+1}^{\prime })^{\prime }$ in the
model $\mathcal{H}_{k,r_{0}}$; with coefficients $\psi _{1}$ and $\psi _{2}$%
, respectively. By the same quadratic approximation of the log-likelihood,%
\begin{equation}
\mathrm{LR}(\mathcal{H}_{k,r_{0}}|\mathcal{H}_{K,r_{0}})=\tilde{\psi}%
_{2}^{(K,r_{0})\prime }\sum_{t=1}^{T}\mathbb{W}_{2\cdot 10,t}^{0}\mathbb{W}%
_{2\cdot 10,t}^{0\prime }\tilde{\psi}_{2}^{(K,r_{0})}+o_{p}(1),  \label{LR0}
\end{equation}%
with%
\begin{equation*}
\tilde{\psi}_{2}^{(K,r_{0})}=\psi _{2,0}+\left( \sum_{t=1}^{T}\mathbb{W}%
_{2\cdot 10,t}^{0}\mathbb{W}_{2\cdot 10,t}^{0\prime }\right)
^{-1}\sum_{t=1}^{T}\mathbb{W}_{2\cdot 10,t}^{0}z_{t}+o_{p}(T^{-1/2}),
\end{equation*}%
and where%
\begin{equation}
\mathbb{W}_{2\cdot 10,t}^{0}=\mathbb{W}_{2t}-\sum_{t=1}^{T}\mathbb{W}_{2t}(%
\mathbb{W}_{0t}^{0\prime },\mathbb{W}_{1t}^{\prime })\left( \sum_{t=1}^{T}%
\left[ 
\begin{array}{cc}
\mathbb{W}_{0t}^{0}\mathbb{W}_{0t}^{0\prime } & \mathbb{W}_{0t}^{0}\mathbb{W}%
_{1t}^{\prime } \\ 
\mathbb{W}_{1t}\mathbb{W}_{0t}^{0\prime } & \mathbb{W}_{1t}\mathbb{W}%
_{1t}^{\prime }%
\end{array}%
\right] \right) ^{-1}\left( 
\begin{array}{c}
\mathbb{W}_{0t}^{0} \\ 
\mathbb{W}_{1t}%
\end{array}%
\right) ,  \label{resid}
\end{equation}%
the least-squares residual of a regression of $\mathbb{W}_{2t}$ on $\mathbb{W%
}_{0t}^{0}$ and $\mathbb{W}_{1t}$. By the same derivations, an approximation
analogous to (\ref{LR0}) applies to $\mathrm{LR}(\mathcal{H}_{k,r}|\mathcal{H%
}_{K,r})$ for $r>r_{0}$, but with $\tilde{\psi}_{2}^{(K,r_{0})}$ replaced by 
$\tilde{\psi}_{2}^{(K,r)}$, and $\mathbb{W}_{2\cdot 10,t}^{0}$ replaced by $%
\mathbb{W}_{2\cdot 10,t}$, which in turn is defined by (\ref{resid}) with $%
\mathbb{W}_{0t}^{0}$ replaced by $\mathbb{W}_{0t}=G^{\prime }(X_{t-1}\otimes
\sigma _{t}^{-1})$. This leads to the following result:%
\begin{eqnarray}
\mathrm{LR}(\mathcal{H}_{k,r_{0}}|\mathcal{H}_{K,r_{0}})-\mathrm{LR}(%
\mathcal{H}_{k,r}|\mathcal{H}_{K,r}) &=&\tilde{\psi}_{2}^{(K,r_{0})\prime
}\sum_{t=1}^{T}\mathbb{W}_{2\cdot 10,t}^{0}\mathbb{W}_{2\cdot 10,t}^{0\prime
}\tilde{\psi}_{2}^{(K,r_{0})}  \notag \\
&&-\tilde{\psi}_{2}^{(K,r)\prime }\sum_{t=1}^{T}\mathbb{W}_{2\cdot 10,t}%
\mathbb{W}_{2\cdot 10,t}^{\prime }\tilde{\psi}_{2}^{(K,r)}+o_{p}(1)  \notag
\\
&=&\psi _{2,0}^{\prime }\sum_{t=1}^{T}\left( \mathbb{W}_{2\cdot 10,t}^{0}%
\mathbb{W}_{2\cdot 10,t}^{0\prime }-\mathbb{W}_{2\cdot 10,t}\mathbb{W}%
_{2\cdot 10,t}^{\prime }\right) \psi _{2,0}  \notag \\
&&+2\psi _{2,0}^{\prime }\sum_{t=1}^{T}\left( \mathbb{W}_{2\cdot 10,t}^{0}-%
\mathbb{W}_{2\cdot 10,t}\right) z_{t}  \notag \\
&&+\sum_{t=1}^{T}z_{t}^{\prime }\mathbb{W}_{2\cdot 10,t}^{0\prime }\left(
\sum_{t=1}^{T}\mathbb{W}_{2\cdot 10,t}^{0}\mathbb{W}_{2\cdot 10,t}^{0\prime
}\right) ^{-1}\sum_{t=1}^{T}\mathbb{W}_{2\cdot 10,t}^{0}z_{t}  \notag \\
&&-\sum_{t=1}^{T}z_{t}^{\prime }\mathbb{W}_{2\cdot 10,t}^{\prime }\left(
\sum_{t=1}^{T}\mathbb{W}_{2\cdot 10,t}\mathbb{W}_{2\cdot 10,t}^{\prime
}\right) ^{-1}\sum_{t=1}^{T}\mathbb{W}_{2\cdot 10,t}z_{t}  \notag \\
&&+o_{p}(1).  \label{LRdif}
\end{eqnarray}%
The third and fourth terms in the final right-hand side expression are $%
O_{p}(1)$, since they represent essentially the two likelihood ratio
statistics under the null hypothesis $\psi _{2}=0$. We will now analyse the
first two terms.

Because $\mathcal{H}_{K,r_{0}}$ is nested in $\mathcal{H}_{K,r}$, it follows
that the column space of $G_{0}$ is a subset of the column space of $G$.
Without loss of generality (after suitable rotation), we may write $%
G=[G_{0}:G^{\ast }]$ for some matrix $G^{\ast }$, orthogonal to $G_{0}$.
Using standard derivations involving projection matrices, this leads to 
\begin{equation*}
\mathbb{W}_{2\cdot 10,t}^{0}-\mathbb{W}_{2\cdot 10,t}=\sum_{t=1}^{T}\mathbb{W%
}_{2t}\mathbb{W}_{0t}^{\ast \prime }\left( \sum_{t=1}^{T}\mathbb{W}%
_{0t}^{\ast }\mathbb{W}_{0t}^{\ast \prime }\right) ^{-1}\mathbb{W}%
_{0t}^{\ast },
\end{equation*}%
with $\mathbb{W}_{0t}^{\ast }$ the least-squares residual of a regression of 
$G^{\ast \prime }(X_{t-1}\otimes \sigma _{t}^{-1})$ on $\mathbb{W}_{0t}^{0}$
and $\mathbb{W}_{1t}$; and%
\begin{equation*}
\sum_{t=1}^{T}\left( \mathbb{W}_{2\cdot 10,t}^{0}\mathbb{W}_{2\cdot
10,t}^{0\prime }-\mathbb{W}_{2\cdot 10,t}\mathbb{W}_{2\cdot 10,t}\right)
=\sum_{t=1}^{T}\mathbb{W}_{2t}\mathbb{W}_{0t}^{\ast \prime }\left(
\sum_{t=1}^{T}\mathbb{W}_{0t}^{\ast }\mathbb{W}_{0t}^{\ast \prime }\right)
^{-1}\mathbb{W}_{0t}^{\ast }\mathbb{W}_{2t}^{\prime }.
\end{equation*}%
It can be shown that $G^{\ast \prime }(X_{t-1}\otimes \sigma _{t}^{-1})$
selects $\mathrm{I}(1)$ linear combinations from $X_{t-1}$, which implies%
\begin{equation*}
\sum_{t=1}^{T}\mathbb{W}_{0t}^{\ast }\mathbb{W}_{0t}^{\ast \prime
}=O_{p}(T^{2}),\qquad \sum_{t=1}^{T}\mathbb{W}_{0t}^{\ast }\mathbb{W}%
_{2t}^{\prime }=O_{p}(T),\qquad \sum_{t=1}^{T}\mathbb{W}_{0t}^{\ast
}z_{t}=O_{p}(T),
\end{equation*}%
and substituting this in (\ref{LRdif}) leads to $\mathrm{LR}(\mathcal{H}%
_{k,r_{0}}|\mathcal{H}_{K,r_{0}})-\mathrm{LR}(\mathcal{H}_{k,r}|\mathcal{H}%
_{K,r})=O_{p}(1)$. Hence, because $\pi (k,r)-\pi (k,r_{0})>0$, it follows
that 
\begin{equation*}
\Pr \left( \mathrm{ALS}\text{-}\mathrm{IC}(k,r)-\mathrm{ALS}\text{-}\mathrm{%
IC}(k,r_{0})>0\right) \rightarrow 1
\end{equation*}%
if $c_{T}\rightarrow \infty $.

For $k<k_{0},r<r_{0}$, the proof follows from a combination of ingredients
from the previous two cases: now

\begin{equation*}
\mathrm{LR}(\mathcal{H}_{k,r}|\mathcal{H}_{k,r_{0}})=\mathrm{LR}(\mathcal{H}%
_{K,r}|\mathcal{H}_{K,r_{0}})+\mathrm{LR}(\mathcal{H}_{k,r}|\mathcal{H}%
_{K,r})-\mathrm{LR}(\mathcal{H}_{k,r_{0}}|\mathcal{H}_{K,r_{0}}).
\end{equation*}%
The first right-hand side term will diverge at the rate $O_{p}(T)$,
analogous to the result for $k\geq k_{0}$, $r<r_{0}$; and the final two
terms together will be $O_{p}(1)$ as in the case $k<k_{0}$, $r\geq r_{0}$.
This again leads to the required result. \hfill $\square $

\bigskip

\noindent \textbf{Proof of Lemma 2.} As in the proof of Lemma 1, we start
with expressing the ALS-IC difference in terms of likelihood ratio
statistics and $\pi (k,r)$. For $k_{0}<k\leq K$,%
\begin{equation*}
\mathrm{ALS}\text{-}\mathrm{IC}(k,r_{0})-\mathrm{ALS}\text{-}\mathrm{IC}%
(k_{0},r_{0})=-\mathrm{LR}(\mathcal{H}_{k_{0},r_{0}}|\mathcal{H}%
_{k,r_{0}})+c_{T}\left[ \pi (k,r_{0})-\pi (k_{0},r_{0})\right] .
\end{equation*}%
The first right-hand side term is an LR test statistic for a true null
hypothesis in a well-specified model, and hence $O_{p}(1)$. Because $\pi
(k,r_{0})-\pi (k_{0},r_{0})>0$, the ALS-IC diverges provided $%
c_{T}\rightarrow \infty $, which proves part (\emph{i}).

For $0<k<k_{0}$, 
\begin{equation*}
\mathrm{ALS}\text{-}\mathrm{IC}(k,r_{0})-\mathrm{ALS}\text{-}\mathrm{IC}%
(k_{0},r_{0})=\mathrm{LR}(\mathcal{H}_{k,r_{0}}|\mathcal{H}%
_{k_{0},r_{0}})+c_{T}\left[ \pi (k,r_{0})-\pi (k_{0},r_{0})\right] .
\end{equation*}%
The first right-hand side term is an LR statistic for a false null
hypothesis in a well-specified model, and hence will diverge at the rate $%
O_{p}(T)$; see the proof of Lemma 1, case $k<k_{0}$, $r>r_{0}$. Since $\pi
(k,r_{0})-\pi (k_{0},r_{0})<0$ in thise case, the ALS-IC diverges provided $%
c_{T}=o(T)$, which proves part (\emph{ii}). \hfill $\square $

\bigskip

\noindent \textbf{Proof of Theorem 1. } The theorem is a direct extension of
Theorem 1 of Cavaliere \emph{et al}.\ (2018) to the case of ALS-based
information criteria. Making use of Lemmas 1 and 2, the line of the proof is
exactly the same as in their proof. \hfill $\square $

\bigskip

\noindent \textbf{Proof of Lemma 3. }The proof is analogous to the proof of
Lemma 2; the difference is that the true cointegrating rank $r_{0}$ in Lemma
2 has been replaced here by $p\geq r_{0}$. Therefore, the LR test statistics
are now for a true or false null hypothesis in an over-specified model; but
this does not affect the divergence rates, hence the same results obtain.
\hfill $\square $

\bigskip

\noindent \textbf{Proof of Theorem 2. }It follows from Boswijk and Zu
(2022), Theorem 3, that when using the true lag length $k_{0}$, the
bootstrap PLR-tests have correct size and are consistent, i.e., for the
chosen significance level $\eta $, and as $T\rightarrow \infty $, 
\begin{eqnarray*}
\Pr \left( \hat{r}^{\ast }(k_{0})<r\right)  &\rightarrow &0, \\
\Pr \left( \hat{r}^{\ast }(k_{0})=r_{0}\right)  &\rightarrow &1-\eta .
\end{eqnarray*}%
Together with Lemma 3, this implies the result of Theorem 2,
analogously to the proof of Theorem 3 of Cavaliere \emph{et al}.\ (2018).
\hfill $\square $

\bigskip

\noindent \textbf{Proof of Theorem 3. }The theorem is a direct extension of
Theorem 2 of Cavaliere \emph{et al}.\ (2018) to the case of ALS-based
information criteria. Making use of Lemmas 1--3, the line of the proof is
exactly the same as in their proof. \hfill $\square $

\pagebreak

\newpage \newgeometry{left=10mm,right=10mm,top=10mm,bottom=10mm} %
\thispagestyle{empty}

\begin{landscape}

\begin{table}[htbp]
TABLE 1: Joint procedures for determining the co-integration rank and the lag length. VAR(2) model with rank $r_0 = 0, 1, 2$ \\
\tiny{
\begin{tabular}{rrrrrrrrrrrrrr|rrrrrrrrrrrr}
\hline \hline
    {\bf } &            &                                                                                                   \multicolumn{ 12}{c}{Co-integration rank determination} &                                                                                                           \multicolumn{ 12}{|c}{Lag length determination} \\

           &            & \multicolumn{ 3}{c}{ALS-HQC($k,r$)} & \multicolumn{ 3}{c}{ALS-BIC($k,r$)} &     \multicolumn{ 3}{c}{HQC($k,r$)} &     \multicolumn{ 3}{c}{BIC($k,r$)} & \multicolumn{ 3}{|c}{ALS-HQC($k,r$)} & \multicolumn{ 3}{c}{ALS-BIC($k,r$)} &     \multicolumn{ 3}{c}{HQC($k,r$)} &     \multicolumn{ 3}{c}{BIC($k,r$)} \\
\hline \hline
  $r_0=0$ &            &                                                                                                                                                                                                                                                            \multicolumn{ 24}{c}{Autoregressive Stochastic Volatility} \\

   {$\gamma$} &          T &   $r = 0$ &          1 &          2 &   $r = 0$ &          1 &          2 &   $r = 0$ &          1 &          2 &   $r = 0$ &          1 &          2 &   $k = 1$ &          2 &        3,4 &   $k = 1$ &          2 &        3,4 &   $k = 1$ &          2 &        3,4 &   $k = 1$ &          2 &        3,4 \\
\hline
       0.0 &         50 & {\bf 65.6} &       28.1 &        6.3 & {\bf 81.4} &       14.8 &        3.8 & {\bf 72.0} &       25.8 &        2.2 & {\bf 89.7} &       10.0 &        0.3 & {\bf 75.9} &       10.9 &       13.2 & {\bf 88.4} &        6.0 &        5.6 & {\bf 78.8} &       12.3 &        8.9 & {\bf 92.2} &        5.7 &        2.1 \\
\hline \hline
\end{tabular} 

}
\end{table}

\end{landscape}

\newpage \thispagestyle{empty}

\begin{landscape}

\begin{table}[htbp]
\scriptsize{TABLE 2: Sequential information criteria-based procedures for determining the lag length (Step I) and the co-integration rank (Step II). VAR(2) model with rank $r_0 = 0, 1, 2$ (AR stochastic volatility)} \\
\tiny{
\begin{tabular}{rrrrrrrrrrrrrrr|rrrrrrrrrrrr}
\hline \hline
           &            &            &    \multicolumn{ 3}{c}{ALS-HQC} &    \multicolumn{ 3}{c}{ALS-BIC} &       \multicolumn{ 3}{|c}{HQC} &        \multicolumn{ 3}{c}{BIC} &   \multicolumn{ 3}{|c}{ALS-HQC} &    \multicolumn{ 3}{c}{ALS-BIC} &       \multicolumn{ 3}{|c}{HQC} &        \multicolumn{ 3}{c}{BIC} \\
\hline
    {\bf } &            &            &                                                                                                                   \multicolumn{ 12}{c}{{$\gamma=0$}} &                                                                                                                \multicolumn{ 12}{|c}{{$\gamma=0.1$}} \\
\hline
\multicolumn{ 2}{l}{Step I}            &          T &   $k = 1$ &          2 &        3,4 &   $k = 1$ &          2 &        3,4 &   $k = 1$ &          2 &        3,4 &   $k = 1$ &          2 &        3,4 &   $k = 1$ &          2 &        3,4 &   $k = 1$ &          2 &        3,4 &   $k = 1$ &          2 &        3,4 &   $k = 1$ &          2 &        3,4 \\
\hline
           &            &         50 & {\bf 75.6} &       12.7 &       11.7 & {\bf 86.9} &        5.8 &        7.3 & {\bf 79.2} &       11.4 &        9.4 & {\bf 92.5} &        4.8 &        2.7 &       68.6 & {\bf 16.8} &       14.6 &       84.1 &  {\bf 8.7} &        7.2 &       70.7 & {\bf 18.8} &       10.5 &       87.9 &  {\bf 9.3} &        2.8 \\

           &            &        100 & {\bf 85.5} &        6.7 &        7.8 & {\bf 93.5} &        2.1 &        4.4 & {\bf 74.1} &       14.4 &       11.5 & {\bf 89.8} &        7.3 &        2.9 &       78.4 & {\bf 13.5} &        8.1 &       88.3 &  {\bf 6.0} &        5.7 &       64.4 & {\bf 23.6} &       12.0 &       84.1 & {\bf 13.2} &        2.7 \\
\hline
 \hline
\end{tabular}  
}
\end{table}

\end{landscape}

\newpage \thispagestyle{empty}

\begin{landscape}

\begin{table}[htbp]
\scriptsize{TABLE 3: Sequential information criteria-based procedures for determining the lag length (Step I) and the co-integration rank (Step II). VAR(2) model with rank $r_0 = 0, 1, 2$ (Single volatility break)} \\
\tiny{
\begin{tabular}{rrrrrrrrrrrrrrr|rrrrrrrrrrrr}
\hline \hline
           &            &            &    \multicolumn{ 3}{c}{ALS-HQC} &    \multicolumn{ 3}{c}{ALS-BIC} &       \multicolumn{ 3}{|c}{HQC} &        \multicolumn{ 3}{c}{BIC} &   \multicolumn{ 3}{|c}{ALS-HQC} &    \multicolumn{ 3}{c}{ALS-BIC} &       \multicolumn{ 3}{|c}{HQC} &        \multicolumn{ 3}{c}{BIC} \\
\hline
    {\bf } &            &            &                                                                                                                   \multicolumn{ 12}{c}{{$\gamma=0$}} &                                                                                                                \multicolumn{ 12}{|c}{{$\gamma=0.1$}} \\
\hline
\multicolumn{ 2}{l}{Step I}            &          T &   $k = 1$ &          2 &        3,4 &   $k = 1$ &          2 &        3,4 &   $k = 1$ &          2 &        3,4 &   $k = 1$ &          2 &        3,4 &   $k = 1$ &          2 &        3,4 &   $k = 1$ &          2 &        3,4 &   $k = 1$ &          2 &        3,4 &   $k = 1$ &          2 &        3,4 \\
\hline
           &            &         50 & {\bf 68.5} &       16.5 &       15.0 & {\bf 86.0} &        8.6 &        5.4 & {\bf 61.6} &       18.1 &       20.3 & {\bf 87.3} &        9.2 &        3.5 &       62.7 & {\bf 18.1} &       19.2 &       81.4 & {\bf 12.2} &        6.4 &       54.5 & {\bf 20.9} &       24.6 &       82.2 & {\bf 13.1} &        4.7 \\

           &            &        100 & {\bf 91.0} &        7.0 &        2.0 & {\bf 97.4} &        1.9 &        0.7 & {\bf 69.9} &       15.4 &       14.7 & {\bf 92.7} &        6.4 &        0.9 &       81.4 & {\bf 14.7} &        3.9 &       94.7 &  {\bf 5.0} &        0.3 &       62.0 & {\bf 21.1} &       16.9 &       87.1 & {\bf 10.5} &        2.4 \\
\hline \hline
\end{tabular}  
}
\end{table}

\end{landscape}

\newpage \thispagestyle{empty}

\begin{landscape}

\begin{table}[htbp]
\footnotesize{TABLE 4: Sequential bootstrap-based procedures for determining the lag length (Step I) and the co-integration rank (Step II). VAR(2) model with rank $r_0 = 0, 1, 2$} (AR Stochastic Volatility) \\
\tiny{
\begin{tabular}{rrrrrrrrrrr|rrrrrrrrr|rrr}
\hline
    {\bf } &            &                                                  \multicolumn{ 9}{c}{Co-integration rank determination ($r_0=0$)} &                                                  \multicolumn{ 9}{c}{Co-integration rank determination ($r_0=1$)} & \multicolumn{ 3}{|c}{Lag length determination} \\
\hline
  $k=k_0$ &            &     \multicolumn{ 3}{c}{PLR-WB} &      \multicolumn{ 3}{c}{ALR-VB} &     \multicolumn{ 3}{c}{ALR-WB} &    \multicolumn{ 3}{|c}{PLR-WB} &      \multicolumn{ 3}{c}{ALR-VB} &     \multicolumn{ 3}{c}{ALR-WB} &            \multicolumn{ 3}{|c}{BIC} \\

 $\gamma$ &          $T$ &   $r = 0$ &          1 &          2 &   $r = 0$ &          1 &          2 &   $r = 0$ &          1 &          2 &   $r = 0$ &          1 &          2 &   $r = 0$ &          1 &          2 &   $r = 0$ &          1 &          2 &   $k = 1$ &          2 &        3,4 \\
\hline
    0.0   & 50    & \textbf{93.0} & 6.5   & 0.5   & \textbf{79.5} & 15.9  & 4.6   & \textbf{82.4} & 15.0  & \multicolumn{1}{r|}{2.6} & 36.8  & \textbf{58.6} & 4.5   & 18.9  & \textbf{71.6} & 9.5   & 19.4  & \textbf{70.5} & 10.1  & {-} & {-} & {-} \\
          & 100   & \textbf{94.6} & 4.8   & 0.6   & \textbf{82.8} & 13.5  & 3.7   & \textbf{82.0} & 15.1  & \multicolumn{1}{r|}{2.9} & 6.3   & \textbf{88.3} & 5.4   & 0.8   & \textbf{87.7} & 11.5  & 3.0   & \textbf{85.1} & 11.9  & {-} & {-} & {-} \\
    0.1   & 50    & \textbf{93.4} & 6.0   & 0.6   & \textbf{81.9} & 14.2  & 3.9   & \textbf{80.2} & 16.5  & \multicolumn{1}{r|}{3.3} & 44.7  & \textbf{50.5} & 4.8   & 25.8  & \textbf{65.6} & 8.5   & 24.9  & \textbf{66.0} & 9.1   & {-} & {-} & {-} \\
          & 100   & \textbf{93.5} & 5.8   & 0.7   & \textbf{87.1} & 10.2  & 2.7   & \textbf{80.2} & 16.9  & \multicolumn{1}{r|}{2.9} & 8.5   & \textbf{86.2} & 5.4   & 3.0   & \textbf{87.9} & 9.1   & 1.5   & \textbf{86.0} & 12.5  & {-} & {-} & {-} \\
\end{tabular}  

}
\end{table}

\end{landscape}

\newpage \thispagestyle{empty}

\begin{landscape}

\begin{table}[htbp]
\footnotesize{TABLE 5: Sequential bootstrap-based procedures for determining the lag length (Step I) and the co-integration rank (Step II). VAR(2) model with rank $r_0 = 0, 1, 2$} (Single Volatility Break) \\
\tiny{
\begin{tabular}{rrrrrrrrrrr|rrrrrrrrr|rrr}
\hline
    {\bf } &            &                                                  \multicolumn{ 9}{c}{Co-integration rank determination ($r_0=0$)} &                                                  \multicolumn{ 9}{c}{Co-integration rank determination ($r_0=1$)} & \multicolumn{ 3}{|c}{Lag length determination} \\
\hline
  $k=k_0$ &            &     \multicolumn{ 3}{c}{PLR-WB} &      \multicolumn{ 3}{c}{ALR-VB} &     \multicolumn{ 3}{c}{ALR-WB} &    \multicolumn{ 3}{|c}{PLR-WB} &      \multicolumn{ 3}{c}{ALR-VB} &     \multicolumn{ 3}{c}{ALR-WB} &            \multicolumn{ 3}{|c}{BIC} \\

 $\gamma$ &          T &   $r = 0$ &          1 &          2 &   $r = 0$ &          1 &          2 &   $r = 0$ &          1 &          2 &   $r = 0$ &          1 &          2 &   $r = 0$ &          1 &          2 &   $r = 0$ &          1 &          2 &   $k = 1$ &          2 &        3,4 \\
\hline
       0.0 &         50 & {\bf 93.5} &        5.8 &        0.7 & {\bf 86.2} &       11.6 &        2.2 & {\bf 88.9} &        8.9 &        2.2 &       45.9 & {\bf 48.6} &        5.5 &       15.4 & {\bf 77.3} &        7.3 &       19.0 & {\bf 71.8} &        9.2 &          - &          - &          - \\

           &        100 & {\bf 94.5} &        4.9 &        0.6 & {\bf 92.2} &        6.9 &        1.0 & {\bf 91.9} &        7.0 &        1.2 &        6.0 & {\bf 88.0} &        6.0 &        0.1 & {\bf 94.0} &        5.9 &        0.1 & {\bf 93.0} &        6.9 &          - &          - &          - \\

       0.1 &         50 & {\bf 92.6} &        6.8 &        0.6 & {\bf 84.5} &       14.2 &        1.3 & {\bf 88.3} &       10.1 &        1.6 &       54.9 & {\bf 39.9} &        5.2 &       24.3 & {\bf 68.5} &        7.2 &       29.9 & {\bf 61.9} &        8.2 &          - &          - &          - \\
\end{tabular}  

}
\end{table}

\end{landscape}

\newpage \thispagestyle{empty}

\begin{landscape}

\begin{table}[htbp]
\footnotesize{TABLE 6: Sequential joint procedures for determining the co-integration rank and the lag length. VAR(2) model with rank $r_0 = 0, 1, 2$ (Homoskedastic DGP)} \\
\tiny{
\begin{tabular}{rrrrrrrrrrrrrr|rrrrrrrrrrrr}
\hline \hline
    {\bf } &            &                                                                                                   \multicolumn{ 12}{c}{Co-integration rank determination} &                                                                                                           \multicolumn{ 12}{|c}{Lag length determination} \\

           &            & \multicolumn{ 3}{c}{ALS-HQC($k,r$)} & \multicolumn{ 3}{c}{ALS-BIC($k,r$)} &     \multicolumn{ 3}{c}{HQC($k,r$)} &     \multicolumn{ 3}{c}{BIC($k,r$)} & \multicolumn{ 3}{|c}{ALS-HQC($k,r$)} & \multicolumn{ 3}{c}{ALS-BIC($k,r$)} &     \multicolumn{ 3}{c}{HQC($k,r$)} &     \multicolumn{ 3}{c}{BIC($k,r$)} \\
\hline \hline
     $r_0=0$ &            &             \multicolumn{24}{c}{} \\

   {$\gamma$} &          T &      r = 0 &          1 &          2 &      r = 0 &          1 &          2 &      r = 0 &          1 &          2 &      r = 0 &          1 &          2 &      k = 1 &          2 &        3,4 &      k = 1 &          2 &        3,4 &      k = 1 &          2 &        3,4 &      k = 1 &          2 &        3,4 \\
\hline
       0.0 &         50 & {\bf 69.3} &       27.3 &        3.4 & {\bf 86.9} &       11.9 &        1.2 & {\bf 82.9} &       16.3 &        0.8 & {\bf 95.9} &        3.9 &        0.2 & {\bf 89.3} &        6.3 &        4.4 & {\bf 96.6} &        2.3 &        1.1 & {\bf 96.5} &        2.6 &        0.9 & {\bf 99.8} &        0.2 &        0.0 \\

           &        100 & {\bf 84.3} &       14.6 &        1.1 & {\bf 96.2} &        3.5 &        0.3 & {\bf 89.1} &       10.3 &        0.6 & {\bf 98.5} &        1.4 &        0.1 & {\bf 97.8} &        2.0 &        0.2 & {\bf 99.6} &        0.4 &        0.0 & {\bf 99.1} &        0.9 &        0.0 & {\bf 99.7} &        0.3 &        0.0 \\
\hline
\end{tabular}

}
\end{table}

\end{landscape}

\pagebreak
\restoregeometry

\begin{table}[htbp]

TABLE 7: Co-integration rank and lag length determination for the term structure of interest rates in the US using standard and adaptive joint information-based procedures, IC($k,r$) and ALS-IC($k,r$).
\center{
\begin{tabular}{rrrrr}
\hline
      &   \ \ \ \ HQC   &  \ \ \ \  BIC   &      ALS-HQC   &    ALS-BIC   \\
\hline
$\tilde{k}$  &     2  &  2 & 2 & 2 \\
$\tilde{r}$  &     3  &  2 & 4  & 2 \\
\hline
\end{tabular}  

\smallskip

\noindent \footnotesize{Note: VAR models are fitted with a restricted constant. The maximum number of lags is $K = 4$.}
}
\end{table}


\pagebreak

\begin{table}[ht]

TABLE 8: Co-integration rank and lag length determination for the term structure of interest rates in the US using standard and adaptive sequential procedures.
\center{
\begin{tabular}{rrrrr}
\hline \hline
           & {\ HQC \ }              & {\ \ BIC \ }             & {ALS-HQC}             &  {ALS-BIC}           \\
\hline
Step I  \ \ \ $\hat{k}$          & 2 \ \ \ \ &  1 \ \ \ \ &  2 \ \ \ \ \   &  2 \ \ \ \ \  \\
\hline \hline
Step II (IC) & & & &  \\
 $\hat{r}_\text{IC}(\hat{k}=1)$ &          4 \ \ \ \ &          2 \ \ \ \ &      5 \ \ \ \ \   &   3 \ \ \ \ \   \\
 $\hat{r}_\text{IC}(\hat{k}=2)$ &          4 \ \ \ \ &          2 \ \ \ \ &      4 \ \ \ \ \   &   2 \ \ \ \ \   \\
\end{tabular}
\begin{tabular}{rrrrrrrr}
\hline \hline
Step II (PLR)         & \multicolumn{ 2}{c}{PLR-WB}            & \multicolumn{ 2}{c}{ALR-VBS}  &      \multicolumn{ 2}{c}{ALR-WBS}          \\
           & $\hat{k}=1$ & $\hat{k}=2$ & $\hat{k}=1$ & $\hat{k}=2$ & $\hat{k}=1$ & $\hat{k}=2$ \\
\hline
   $r=0$  &      0.000 &      0.000 &      0.000 &      0.000 &      0.000 &      0.000   \\

  $r=1$  &      0.000 &      0.000 &      0.000 &      0.000 &      0.000 &      0.000       \\

  $r=2$  &      0.015 &      0.091 &      0.000 &      0.000 &      0.000 &      0.001      \\

  $r=3$  &      0.270 &      0.309 &      0.010 &      0.038 &      0.000 &      0.011     \\

  $r=4$  &      0.914 &      0.806 &      0.082 &      0.172 &      0.035 &      0.124    \\

$\hat{r}^\ast_{(\eta=0.10)}$ & 3 \ \ & 3 \ \  & 5 \ \  & 4 \ \  & 5 \ \  & 4 \ \  \\
$\hat{r}^\ast_{(\eta=0.05)}$ & 3 \ \  & 2 \ \  & 4 \ \  & 4 \ \  & 5 \ \  & 4 \ \  \\
$\hat{r}^\ast_{(\eta=0.01)}$ & 2 \ \  & 2 \ \  & 3  \ \ & 3 \ \  & 4 \ \  & 3 \ \  \\
\hline \hline
\end{tabular}  }

\smallskip

\noindent \footnotesize{Notes: VAR models are fitted with a restricted constant. The maximum number of lags is $K = 4$.

`PLR-WB' denotes the (non-adaptive) wild bootstrap PLR test-based approach; `ALR-VB' denotes the adaptive PLR test based on the volatility bootstrap; `ALR-WB' denotes the adaptive PLR test based on the wild bootstrap.
The number $B$ of bootstrap samples used in the wild bootstrap algorithm is 999.
}
\end{table}

\end{document}